%% file: main.tex
\documentclass[hyphens]{article} 
\usepackage[preprint]{colm2026_conference}

\usepackage{microtype}
\usepackage{hyperref}
\usepackage{url}
\usepackage{booktabs}
\usepackage{lineno}
\usepackage{subcaption}
\usepackage{graphicx}
\usepackage{tcolorbox}
\tcbuselibrary{breakable}
\usepackage{xcolor}
\usepackage{tabularx}
\usepackage{booktabs}
\usepackage{array}
\usepackage{ragged2e}
\newcolumntype{L}[1]{>{\RaggedRight\arraybackslash}p{#1}}
\usepackage{inconsolata}
\usepackage{colortbl}
\usepackage{wrapfig}
\usepackage{amsmath}
\usepackage{multirow}
\usepackage{longtable}

\definecolor{darkblue}{rgb}{0, 0, 0.5}
\hypersetup{colorlinks=true, citecolor=darkblue, linkcolor=darkblue, urlcolor=darkblue}

\definecolor{linegray}{gray}{0.97}
\definecolor{salmonred}{HTML}{F65314}
\definecolor{grassgreen}{HTML}{6aa84f}
\definecolor{headerorange}{HTML}{FF7F00}
\definecolor{contextpurple}{HTML}{8B8BAE}
\definecolor{darkgray}{gray}{0.3}
\definecolor{redhl}{HTML}{FCC6BB}
\definecolor{greenhl}{HTML}{b6d7a8}
\definecolor{yellowhl}{HTML}{FFE599}

\definecolor{darkblue}{RGB}{0,0,139}

\tcbset{
  promptstyle/.style={
    colback=linegray,
    colframe=white,
    boxrule=0pt,
    sharp corners,
    left=6pt,
    right=6pt,
    top=4pt,
    bottom=4pt,
    fontupper=\ttfamily\fontsize{7.5}{9}\selectfont,
    before upper=\raggedright,
    breakable,
  }
}
\newtcolorbox{promptbox}[1][]{promptstyle,#1}

\tcbset{
  on line,
  boxsep=0pt,
  left=1pt,
  right=1pt,
  top=0pt,
  bottom=0pt,
  colframe=red!20!black,
  colback=red!20,
  arc=2pt,
  auto outer arc,
  box align=base
}
\newtcbox{\redhl}{on line, colback=redhl, colframe=redhl, arc=3pt, boxrule=0pt, boxsep=0.5pt, left=2pt, right=2pt, top=1pt, bottom=1pt}
\newtcbox{\greenhl}{on line, colback=greenhl, colframe=greenhl, arc=3pt, boxrule=0pt, boxsep=0.5pt, left=2pt, right=2pt, top=1pt, bottom=1pt}
\newtcbox{\yellowhl}{on line, colback=yellowhl, colframe=yellowhl, arc=3pt, boxrule=0pt, boxsep=0.5pt, left=2pt, right=2pt, top=1pt, bottom=1pt}

\title{From Images to Tasks: Characterizing Multimodal LLM \\Interactions in the Wild}

\author{Jinyi Ye\textsuperscript{1,2}\thanks{Work done during internship at Microsoft IDEAS and Microsoft Research.}, Scott Counts\textsuperscript{2}, Gaurav Verma\textsuperscript{2}, Kate Lytvynets\textsuperscript{3}, Weiwei Yang\textsuperscript{3} \\
\textsuperscript{1}University of Southern California, \textsuperscript{2}Microsoft, \textsuperscript{3}Microsoft Research\\
\texttt{\{jinyiy@usc.edu, counts@microsoft.com, gauravverma@microsoft.com\}}
}

\begin{document}

\ifcolmsubmission
\linenumbers
\fi

\maketitle

\begin{abstract}
Multimodal large language models (LLMs) increasingly integrate vision and text, yet how people use them in natural settings remains underexplored.
We seek to answer the question: \textit{when users upload images, what tasks are they trying to accomplish?}
Analyzing over 40,000 de-identified image-upload conversations from Microsoft Copilot, we characterize real-world multimodal use through a hierarchical framework of ten capabilities, spanning perception, cognition, and generation.
First, we characterize the distribution and composition of these capabilities, finding that the majority of image-upload tasks involve multiple capabilities.
Second, we find that multimodal use spans a broader and more diverse task space than text-only interactions, with asymmetric coverage and task classes that rely on cross-modal grounding.
Third, mapping observed capability demand onto 253 existing benchmarks reveals uneven alignment between benchmark coverage and real-world use: benchmarks concentrate on perception and reasoning toward fixed answers, while common workflows involving text, code, and data generation remain comparatively undertested.
We validate our taxonomy and findings on an independent ChatGPT dataset. 
Our results provide a large-scale empirical characterization of what users seek to accomplish with multimodal LLMs and highlight opportunities for benchmark design grounded in observed user demand.
\end{abstract}

\section{Introduction}
\label{sec:intro}

Multimodal interactions, which integrate text inputs with images, documents, audio and videos, offer a more natural and efficient way to engage with conversational AI systems \citep{zhang2024mm, sundar2022multimodal, koh2023grounding}. In typical workflows, people fluidly work across different media: data may be presented in visual plots or tables, websites captured as screenshots, or information conveyed through photos and diagrams. Although most generative AI chatbots now support multimodal inputs \citep{openai2023chatgptcanseehear, team2023gemini, ray2025beyond}, large-scale analyses of user interactions remain focused on the text modality only (e.g., public ChatGPT conversations and analyses \citep{zhao2024wildchat, chatterji2025people}, telemetry analyses of Copilot usage \citep{hoak2025engagement}, occupational task mapping \citep{handa2025economic}, and use-case taxonomies drawn from public discussions \citep{cheng2025realm}). We still lack empirical insight into how often users invoke \textit{multimodal} features in their inputs and which tasks they perform with them in the wild.


In order to characterize multimodal interactions with large language models (LLMs) at the task and workflow levels, we should first ask: What \textit{multimodal capabilities} do users' tasks require when they attach non-text inputs in conversations? Real tasks typically chain capabilities across three layers. \emph{Perception} extracts structured signals from non-text media and aligns them with language \cite{baechler2024screenai, liu2024ocrbench, radford2021learning}. \emph{Cognition} composes these signals with the model's knowledge to reason over interleaved inputs \cite{fu2026mme, yu2024mm, yue2024mmmu}. \emph{Generation} produces the outputs users ultimately request, like text, image, structured documents, or code~\citep{ai2025ming}. A gap remains in mapping real-world, multi-turn use cases to the specific multimodal capabilities and their combinations that these tasks exercise in practice.

Beyond establishing when and how people invoke multimodal features, we also need to know \textit{why} they choose them and \textit{what marginal value} these inputs add over text alone. It remains unclear how multimodal usage relates to text-only usage. Some tasks, such as debugging code or translating short passages, can be represented equally well in text or as screenshots; others are inherently multimodal because success depends on combining visual and textual cues, as when a clinician interprets a medical image alongside notes \citep{corrado2024advancing} or a developer pastes a UI screenshot to generate its HTML/CSS code \citep{agentnative2023screenshots}. Whether and to what extent multimodal and text-only task spaces overlap or diverge is therefore a question to be explored.

Finally, given the taxonomy and distribution of real-world multimodal use, a natural follow-up question is: do existing \emph{multimodal benchmarks} evaluate the capabilities (and combinations) that users actually invoke?
Recent benchmarks (e.g., \citet{fu2026mme}) have evaluated multimodal capabilities spanning perception, reasoning, instruction following, interleaved interactions, and long-context understanding. Their coverage, however, has not been examined against actual usage in the wild.

\begin{figure}[t]
    \centering
    \makebox[\linewidth][c]{%
        \includegraphics[width=1.2\linewidth]{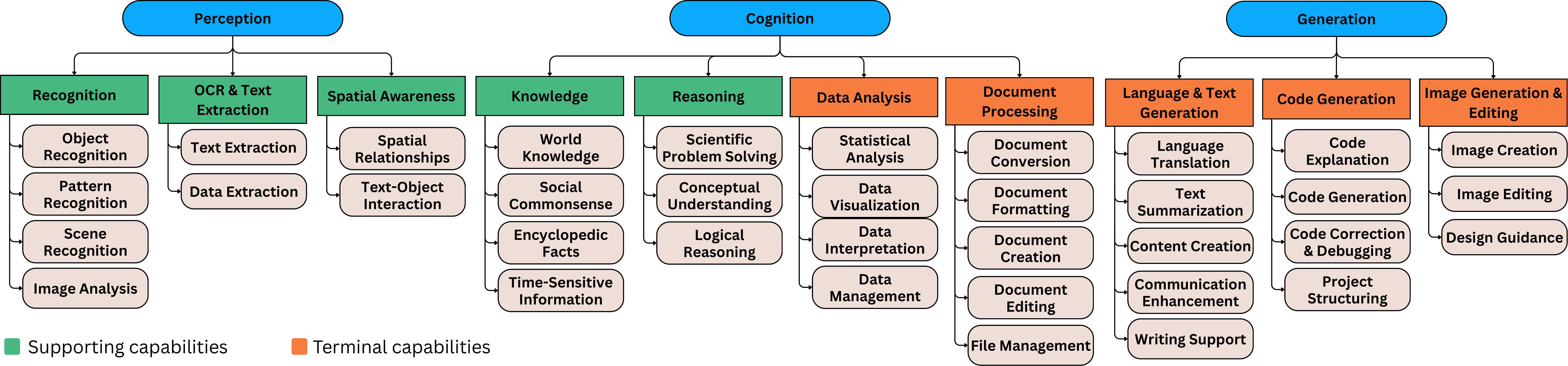}%
    }
    \caption{Hierarchical taxonomy of multimodal capabilities derived from large-scale Copilot and ChatGPT conversations. Ten primary capabilities are organized into perception, cognition and generation, with finer-grained subcategories shown beneath each capability.}
    \label{fig:taxonomy}
\end{figure}

In this study, we bridge these gaps by investigating multimodal interactions with LLMs through a large-scale analysis of image-upload conversations from two widely used AI assistants (Microsoft Copilot and ChatGPT). 
Our contributions are threefold. 
\emph{First}, we derive a hierarchical taxonomy (Figure~\ref{fig:taxonomy}) of multimodal capabilities from real-world interactions and characterize their distributions and combinations.
\emph{Second}, we compare the semantic task spaces of multimodal and text-only interactions to examine the extent to which multimodal usage overlaps with or extends the text-only task space.
\emph{Third}, we use the taxonomy to evaluate the coverage of existing multimodal benchmarks and identify gaps between benchmark design and real-world usage.
Our findings provide empirical basis for designing multimodal benchmarks and conversational interfaces that better reflect how users interact with multimodal LLMs.

\section{Related Works}

\paragraph{Understanding LLM use in the wild.}
Recent work has begun to characterize how people use large language models in natural settings through large-scale analyses of real-world usage. Prior studies span opt-in public conversation corpora~\citep{zhao2024wildchat}, privacy-preserving analyses of millions of on-platform conversations~\citep{tamkin2024clio,chatterji2025people}, occupational and economic tasks performed with LLMs~\citep{handa2025economic}, and use cases mined from public discussions and news reports~\citep{cheng2025realm}. Recent work has extended such analyses to multimodal interactions through crowdsourced VLM evaluation platforms, including WildVision and VisionArena~\citep{lu2024wildvision,chou2025visionarena}. We complement these studies by analyzing naturalistic image-upload interactions from general-purpose production assistants to characterize what users seek to accomplish with multimodal LLMs.

\paragraph{Evaluating multimodal capabilities.}
Multimodal LLMs are commonly evaluated with benchmarks targeting capabilities such as visual perception, OCR, spatial reasoning, knowledge reasoning, and code generation~\citep{fu2026mme, liu2024mmbench, li2024seed, si2025design2code}. Broad benchmark suites aggregate many such tasks into unified evaluations~\citep{yue2024mmmu, ying2024mmt, chen2025mega}, while recent work extends evaluation to integrated capabilities~\citep{yu2024mm2}, interleaved multimodal interactions~\citep{xia2025mmie}, and long-context understanding~\citep{wang2024needle}. Surveys document this rapidly expanding benchmark landscape~\citep{li2024survey}. However, benchmark coverage remains largely top-down, reflecting the evolution of vision–language research rather than observed user behavior. Even benchmarks and evaluation platforms motivated by real-world use, including WildVision and VisionArena, are designed around curated task categories, model comparison, or human preference~\citep{lu2024wildvision,chou2025visionarena,chen2025mega,bitton2023visit}. We complement this line of work by using large-scale interactions to characterize which multimodal capabilities and capability combinations users actually invoke, providing an empirical basis for future benchmark design.

\section{Datasets}

\paragraph{Primary dataset.}
Our primary dataset consists of a two-month anonymous random sample of Microsoft Copilot conversations collected between August and September 2024. The sample excludes enterprise, educational, and commercial accounts.

\paragraph{Validation dataset.}
To assess the reliability and generalizability of our taxonomy and empirical findings, we repeat the same analyses on an independent ChatGPT dataset spanning November 2024 and January 2025. Additional statistics on user intents, application domains and validation results are provided in Appendix~\ref{appendix:results}.

\paragraph{Multimodal sessions.}
A \emph{session} is a complete user conversation with the assistant, consisting of one or more turns. A \emph{multimodal session} is any session in which the user uploads at least one image. We focus exclusively on multimodal \emph{inputs}, excluding multimodal outputs such as text-to-image generation. This yields 42,617 multimodal sessions from Microsoft Copilot and 23,413 from ChatGPT, forming the basis of our analysis.

\paragraph{Privacy considerations.} 
All data used in this study are de-identified prior to analysis. Raw conversation transcripts are first processed by automated systems to remove personally identifiable information, including names, contact details, and other sensitive attributes. These sessions are further summarized into concise, privacy-preserving representations, without reproducing original user text or requiring access to raw images. All subsequent stages, including taxonomy generation, labeling, and hierarchical refinement, operate exclusively on these summaries. All processing takes place within secure, access-controlled environments with strict data retention policies. Internal privacy reviews are conducted before and throughout the study. All findings are reported in aggregate form.

\section{Developing a Taxonomy of Multimodal Capabilities}
\label{sec:taxonomy}

Our first research objective is to understand \emph{what multimodal capabilities users invoke when they upload images to LLMs}. Specifically, we ask: (i) What multimodal capabilities do real-world image-upload tasks require? and (ii) How frequently are these capabilities and their combinations invoked?

We conduct our analysis at the session level, treating each conversation as a user \emph{task} supported by one or more uploaded images. We first derive a hierarchical taxonomy of multimodal capabilities from these interactions and then annotate each session with the capabilities it invokes. This enables us to characterize the distribution of individual capabilities and the compositional patterns through which they are combined.

\begin{figure}[t]
    \centering
    \includegraphics[width=0.6\linewidth]{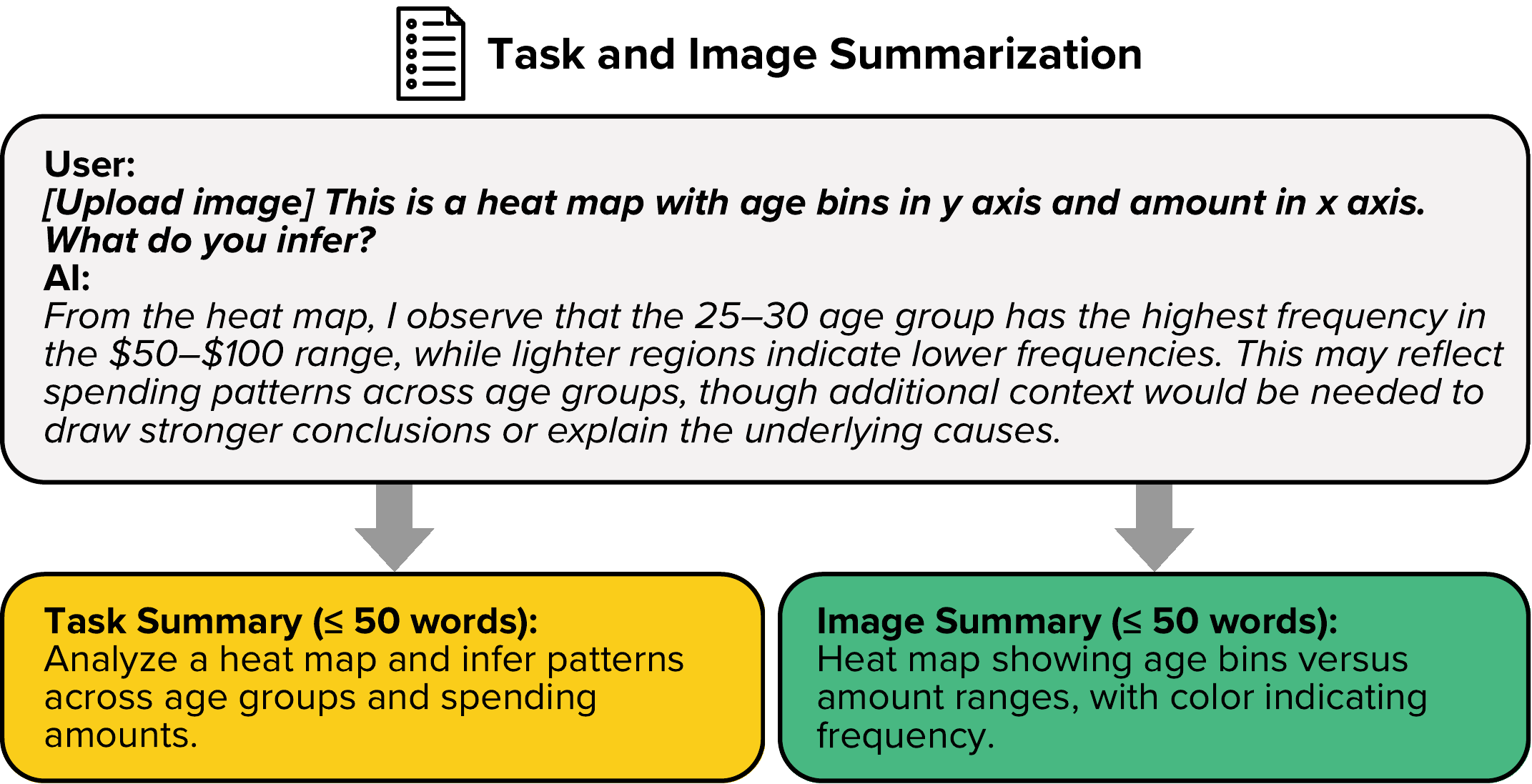}
    \caption{Example task and image summaries extracted from a user–AI conversation.}
    \label{fig:summary_example}
\end{figure}


\begin{table*}[t]
\centering
\caption{Final 10-category taxonomy of multimodal capabilities with descriptions and representative examples.}
\label{tab:taxonomy}
{\fontsize{7.7}{8.8}\selectfont
\begin{tabularx}{\textwidth}{@{}L{2.5cm} c L{4cm} p{5.5cm}@{}}
\toprule
\textbf{Label} & \textbf{Count} & \textbf{Description} & \textbf{Example Task Summaries} \\
\midrule

Recognition & 12,095 & Identify, classify, and describe objects, scenes, actions, and attributes within images. &
\textit{\textbullet\ Identify problems and suggest solutions in a store layout.} \newline
\textit{\textbullet\ Detect objects and count items in a product display.} \newline
\textit{\textbullet\ Classify structural defects in a building image.} \\
\addlinespace[0.4em]

Knowledge & 7,226 & Retrieve, synthesize, and apply factual, commonsense, and contextual information—from both text and visual inputs. &
\textit{\textbullet\ Classify a study as quasi-experimental and seek supporting evidence.} \newline
\textit{\textbullet\ Explain themes of a literary quote with historical context.} \newline
\textit{\textbullet\ Interpret a scientific image and infer material properties.} \\
\addlinespace[0.4em]

OCR \& Text Extraction & 7,136 & Detect, read, and interpret text embedded within images/documents via OCR. &
\textit{\textbullet\ Convert image-based text into readable format for algorithm questions.} \newline
\textit{\textbullet\ Extract names from a scanned list.} \newline
\textit{\textbullet\ Recognize text labels in a form and match them to fields.} \\
\addlinespace[0.4em]

Language \& Text Generation & 3,072 & Produce coherent and context-aware natural-language outputs based on multimodal inputs. &
\textit{\textbullet\ Create a flashcard summarizing a policy document.} \newline
\textit{\textbullet\ Translate a menu from Spanish to English.} \newline
\textit{\textbullet\ Draft an email summary of a chart's main insights.} \\
\addlinespace[0.4em]

Reasoning & 2,653 & Apply structured logic and general knowledge to draw inferences, analyze relationships, and solve complex problems. &
\textit{\textbullet\ Perform long division to find a common factor.} \newline
\textit{\textbullet\ Analyze cause-effect in a science diagram.} \newline
\textit{\textbullet\ Evaluate logical consistency of a sequence of visual steps.} \\
\addlinespace[0.4em]

Data Analysis & 1,510 & Analyze data from charts, graphs, tables, and documents; identify trends and produce structured insights. &
\textit{\textbullet\ Analyze bike rental trends from a Google Sheet.} \newline
\textit{\textbullet\ Summarize numeric values in a budget chart.} \newline
\textit{\textbullet\ Compare sales patterns across two graph images.} \\
\addlinespace[0.4em]

Code Generation & 1,328 & Create, modify, and interpret code across languages, including from visual prompts like screenshots or UI elements. &
\textit{\textbullet\ Solve a shader optimization problem from diagrammed inputs.} \newline
\textit{\textbullet\ Fix syntax errors in a C function snippet.} \newline
\textit{\textbullet\ Generate a Python script to process table data.} \\
\addlinespace[0.4em]

Document Processing & 654 & Ingest and transform content across file types—summarizing, extracting, reformatting, and converting documents. &
\textit{\textbullet\ Merge spreadsheet columns into a summary cell.} \newline
\textit{\textbullet\ Extract tables from a scanned contract.} \newline
\textit{\textbullet\ Reformat a PDF report into editable form.} \\
\addlinespace[0.4em]

Image Generation \& Editing & 617 & Transform or create images via user input—editing, inpainting, style transfer, and generating new visuals. &
\textit{\textbullet\ Add a caption and CTA to a social image.} \newline
\textit{\textbullet\ Remove background and adjust contrast of a product photo.} \newline
\textit{\textbullet\ Generate a sketch from a design prompt.} \\
\addlinespace[0.4em]

Spatial Awareness & 481 & Interpret and reason about positions, relationships, and layouts of objects or text within visual inputs. &
\textit{\textbullet\ Analyze geometric equality of segments in a diagram.} \newline
\textit{\textbullet\ Assess alignment of roads on a map.} \newline
\textit{\textbullet\ Interpret object positions in a flowchart layout.} \\
\bottomrule
\end{tabularx}
}
\end{table*}

We adopt the TnT-LLM framework~\citep{wan2024tnt} to derive a taxonomy of multimodal capabilities and annotate our dataset at scale. TnT-LLM iteratively generates and refines a taxonomy using LLM reasoning over unlabeled data. Building on this framework, we implement the following pipeline using \texttt{gpt-4o-mini} (full prompts are provided in Appendix~\ref{appendix:prompt}).

\paragraph{Summarization.}
To standardize inputs while preserving privacy, we generate two summaries for each session (each limited to 50 words): (i) a \emph{task summary} describing the user's overall objective, and (ii) an \emph{image summary} describing the uploaded image and the requested operation. Because raw images are inaccessible, we adopt an \emph{eyes-off} setting and rely exclusively on these summaries throughout the subsequent analyses. Figure~\ref{fig:summary_example} shows an example.

\paragraph{Taxonomy induction.}
We randomly sample 10,000 sessions for taxonomy induction. Starting from seed categories adapted from prior multimodal benchmarks~\citep{yu2024mm}, we iteratively refine capability names and descriptions over 50 batches of 200 sessions until convergence. The resulting taxonomy contains ten multimodal capability categories (Table~\ref{tab:taxonomy}), with concise descriptions and examples in the main paper and detailed definitions in Appendix~\ref{appendix:description}. 


\paragraph{Dataset annotation.}
Using the induced taxonomy, \texttt{gpt-4o-mini} assigns each session a primary capability and, when applicable, a secondary capability. These annotations enable analyses of both individual capability frequencies and capability combinations across real-world multimodal tasks.

\paragraph{Hierarchical refinement.}
For each of the ten capability categories, we reapply the taxonomy induction procedure to derive finer-grained subcategories, yielding the hierarchical taxonomy shown in Figure~\ref{fig:taxonomy}. The resulting capabilities naturally organize into three functional layers---\emph{perception}, \emph{cognition}, and \emph{generation}---following prior work~\citep{fu2026mme, ai2025ming}. We further distinguish between \emph{supporting capabilities}, which enable downstream tasks, and \emph{terminal capabilities}, which correspond to users' intended outputs (Figure~\ref{fig:taxonomy}).

\paragraph{Validation.}
We evaluate the pipeline along three dimensions: taxonomy robustness across prompt configurations and datasets, agreement between LLM and human annotations ($N=200$), and an \emph{eyes-on} audit verifying that privacy-preserving summaries preserve the capability signal present in the original conversations ($N=200$). Across all evaluations, the taxonomy remains stable, LLM annotations closely match human judgments, and labeling from summaries achieves comparable performance to labeling from raw conversations. Full protocols and quantitative results are reported in Appendix~\ref{appendix:human_validation}.

\section{Results}
\subsection{Capability Distribution Analysis}
\label{sec:distribution}

\begin{table}[t]
\centering
\caption{Top 20 keywords with highest TF–IDF scores by primary capability category. Red = verbs; green = nouns/adjectives.}
\label{tab:tfidf_keywords}
{\fontsize{6}{5}\selectfont
\begin{tabularx}{\textwidth}{@{}c L{2cm} X@{}}
\toprule
\textbf{Id} & \textbf{Label} & \textbf{Top 20 Keywords (TF-IDF)} \\
\midrule
1 & Recognition &
\greenhl{description}, \redhl{analyzes}, \greenhl{detailed}, \greenhl{content}, \redhl{provides}, \redhl{requests}, \greenhl{elements}, \greenhl{point}, \greenhl{recognition}, based, wise, \redhl{asks}, \redhl{identify}, \greenhl{information}, \redhl{analyze}, \redhl{describes}, \redhl{recognizes}, \redhl{questions}, \greenhl{analysis}, \greenhl{details} \\
\addlinespace
2 & Knowledge &
\redhl{asks}, \greenhl{questions}, \redhl{provides}, \greenhl{information}, \redhl{detailed}, \redhl{answers}, \greenhl{related}, \greenhl{specific}, \greenhl{assistant}, \redhl{requests}, llm, \redhl{provided}, \greenhl{steps}, based, including, \redhl{explains}, \greenhl{answer}, \greenhl{explanations}, \greenhl{correct}, \greenhl{summary} \\
\addlinespace
3 & OCR \& Text Extraction &
\greenhl{text}, \redhl{extract}, \greenhl{ocr}, \redhl{containing}, \redhl{extracts}, \greenhl{questions}, \redhl{extracted}, \greenhl{extraction}, \greenhl{content}, \greenhl{pdf}, \greenhl{document}, \redhl{requests}, \greenhl{recognition}, \redhl{convert}, \redhl{format}, \greenhl{data}, \redhl{answers}, \greenhl{excel}, \greenhl{table}, \greenhl{information} \\
\addlinespace
4 & Spatial Awareness &
\greenhl{diagram}, \greenhl{flowchart}, \greenhl{layout}, \greenhl{graph}, based, \redhl{asks}, \redhl{design}, \redhl{provides}, \greenhl{step}, \greenhl{relationships}, \greenhl{map}, \greenhl{geometric}, \greenhl{flow}, \redhl{analyzes}, \redhl{determine}, \greenhl{specific}, \greenhl{components}, \redhl{detailed}, \redhl{requests}, \redhl{solve} \\
\addlinespace
5 & Language \& Text Generation &
\greenhl{resume}, \greenhl{caption}, \redhl{requests}, \greenhl{description}, \greenhl{letter}, \greenhl{post}, \greenhl{text}, \redhl{provides}, based, \redhl{generate}, \redhl{generates}, \greenhl{email}, \redhl{create}, \greenhl{presentation}, \greenhl{assistant}, \greenhl{content}, llm, \redhl{asks}, \greenhl{job}, \greenhl{cover} \\
\addlinespace
6 & Reasoning &
\greenhl{step}, \redhl{solve}, \greenhl{problem}, \greenhl{mathematical}, \greenhl{equation}, \greenhl{calculations}, \redhl{calculates}, \greenhl{problems}, \greenhl{solution}, \greenhl{equations}, involving, \greenhl{expression}, \redhl{provides}, \redhl{answer}, \redhl{asks}, based, \greenhl{function}, given, \greenhl{solutions}, \greenhl{values} \\
\addlinespace
7 & Code Generation &
\greenhl{code}, \greenhl{html}, \greenhl{css}, \greenhl{python}, \greenhl{java}, \greenhl{program}, using, \redhl{provides}, \greenhl{assistant}, \redhl{requesting}, llm, \redhl{requests}, \redhl{create}, \greenhl{snippet}, based, \redhl{design}, \greenhl{javascript}, \greenhl{react}, \greenhl{function}, \greenhl{layout} \\
\addlinespace
8 & Data Analysis &
\greenhl{data}, \greenhl{analysis}, \greenhl{graph}, \greenhl{excel}, \greenhl{chart}, based, \redhl{analyze}, \greenhl{table}, \greenhl{file}, \greenhl{calculations}, \redhl{provides}, \greenhl{insights}, \greenhl{financial}, \redhl{requests}, \redhl{analyzes}, \redhl{provided}, \redhl{calculates}, \redhl{asks}, \greenhl{assistant}, \greenhl{dataset} \\
\addlinespace
9 & Image Generation \& Editing &
\greenhl{background}, \greenhl{photo}, \greenhl{design}, \greenhl{editing}, \greenhl{tools}, \redhl{remove}, \greenhl{logo}, \redhl{requests}, \redhl{create}, \redhl{requesting}, \greenhl{assistant}, \redhl{request}, llm, \redhl{enhance}, \greenhl{color}, \redhl{change}, \greenhl{quality}, \greenhl{resolution}, \greenhl{enhancement}, \redhl{make} \\
\addlinespace
10 & Document Processing &
\greenhl{pdf}, \greenhl{document}, \greenhl{file}, \greenhl{resume}, \greenhl{powerpoint}, \greenhl{presentation}, \greenhl{word}, \greenhl{content}, \redhl{summarize}, \redhl{review}, \greenhl{summarization}, \greenhl{excel}, \redhl{format}, \greenhl{summary}, \redhl{requests}, \redhl{asks}, \redhl{uploading}, needs, \greenhl{copilot}, \redhl{provides} \\
\bottomrule
\end{tabularx}
}
\end{table}

\begin{figure}[t]
    \centering
    \makebox[\linewidth][c]{%
        \includegraphics[width=1.02\linewidth]{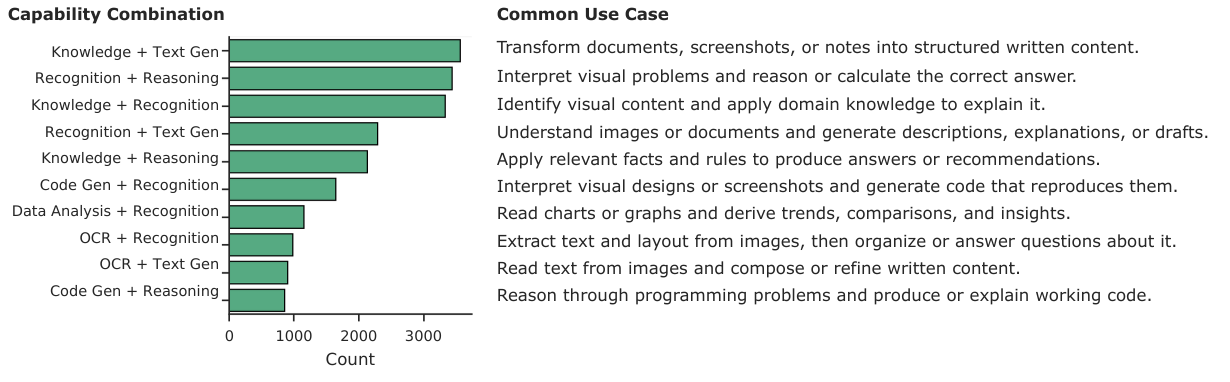}%
    }
    \caption{Top ten multimodal capability combinations in real-world image-upload sessions.}
    \label{fig:capability_combinations}
\end{figure}

Our first research objective is to characterize the multimodal capabilities required by real-world image-upload tasks and how they are composed in practice.
Capability frequencies are computed by aggregating both primary and secondary labels assigned to each session (Table~\ref{tab:taxonomy}). Overall, \textit{Recognition}, \textit{Knowledge}, and \textit{OCR \& Text Extraction} emerge as the three most frequently invoked capabilities in user interactions.

\paragraph{Task charaterization for multimodal capabilities.}
To further characterize representative tasks for each multimodal capability, we analyze task summaries using a TF–IDF vectorizer, following established practices in keyword extraction and text classification \citep{zhang2011comparative}. For each capability, we extract the top 20 keywords with the highest TF–IDF scores, considering both primary and secondary labels. Keywords that occur disproportionately within a category serve as indicators of its content \citep{hulth2006study}. 
We visualize the results in Table~\ref{tab:tfidf_keywords}, with verbs highlighted in green and nouns/adjectives in red. Representative tasks can be interpreted through verb–noun combinations, such as ``provide description'' and ``identify elements'' for \textit{Recognition}, or ``generate resume'' and ``create caption'' for \textit{Language \& Text Generation}. 

\paragraph{Multimodal interactions often involve multiple capabilities.}
We also find that real-world multimodal interactions are inherently compositional: \textbf{74.17\%} of image-upload sessions invoke multiple capabilities rather than a single capability. Figure~\ref{fig:capability_combinations} shows the ten most frequent capability combinations. It reveals that users commonly couple perceptual capabilities (e.g., \textit{Recognition} or \textit{OCR \& Text Extraction}) with downstream cognition (e.g., \textit{Knowledge} or \textit{Reasoning}) and generation (e.g., \textit{Language \& Text Generation} or \textit{Code Generation}). Among these, \textit{Knowledge + Language \& Text Generation} is the single most frequent combination, followed by \textit{Recognition + Reasoning} and \textit{Knowledge + Recognition}. For more context, readers can find representative task examples for the top capability combinations in Appendix~\ref{appendix:combination_examples} (Table~\ref{tab:top_combinations}).


\subsection{Task Space Analysis Between Multimodal and Text-Only Interactions}
\label{sec:taskspace}

Our second research objective is to compare the semantic task spaces of multimodal and text-only interactions. To ensure comparability, we sample text-only sessions of equal size from both Microsoft Copilot and ChatGPT. All sessions are summarized using the same procedure described in Section~\ref{sec:taxonomy}. Because text-only sessions contain no uploaded images, only task summaries are generated. We compute embeddings for all task summaries using \texttt{text-embedding-3-large}, which form the basis for subsequent analyses. We evaluate the relationship between the two task spaces using three complementary analyses: (i) cross-modality perplexity, (ii) out-of-distribution (OOD) detection in embedding space, and (iii) t-SNE visualization.

\begin{wrapfigure}[21]{rt}{0.4\textwidth}
  \vspace{-1.0\baselineskip}
  \centering
  \includegraphics[width=0.9\linewidth]{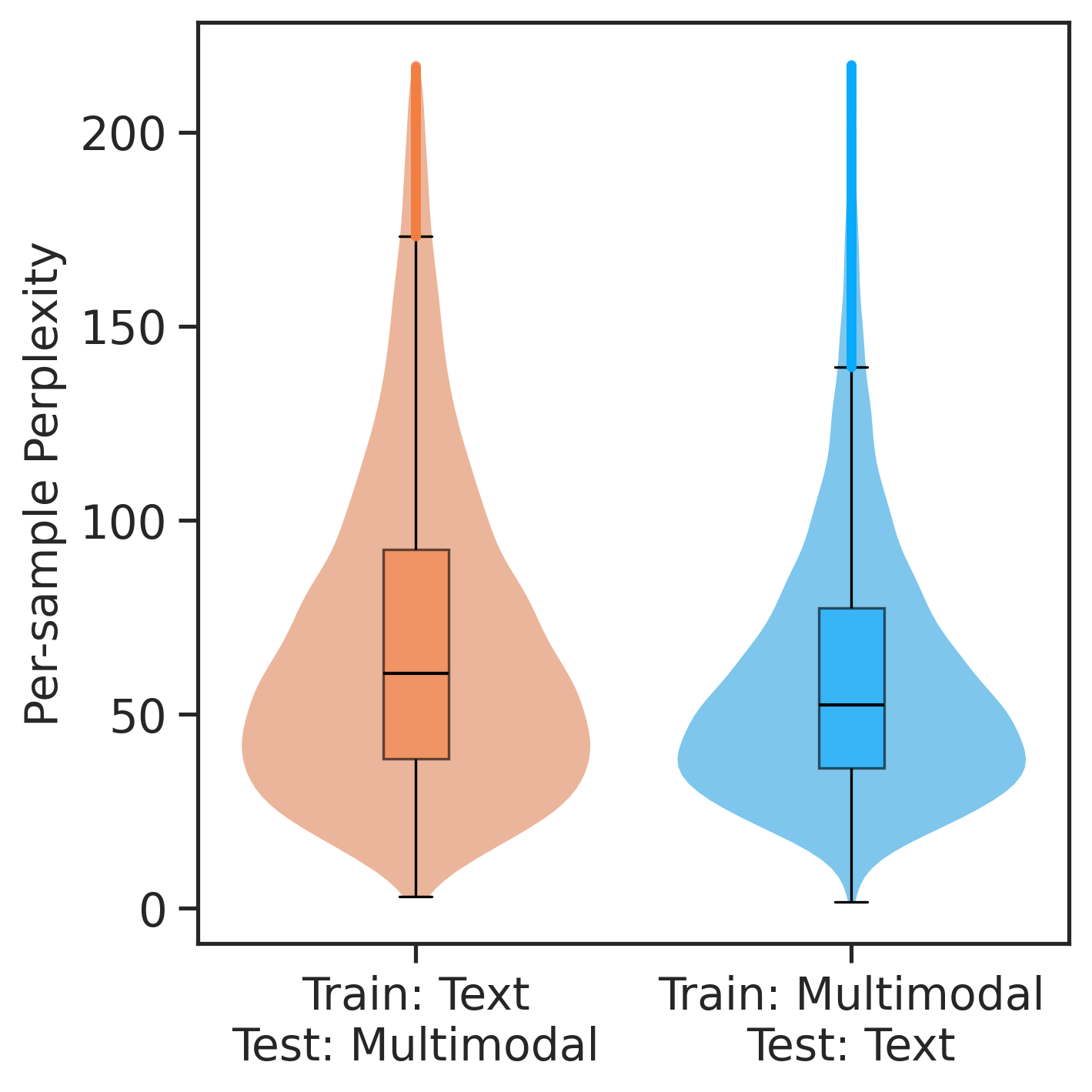}
  \captionsetup{format=plain} 
  \begin{minipage}{0.9\linewidth} 
    \vspace{1.0\baselineskip}
    \caption{Cross-modality comparison of per-sample perplexity between models fine-tuned on text-only and multimodal datasets, evaluated using a fine-tuned Llama-3.1 8B model.}
    \label{fig:perplexity}
  \end{minipage}
\end{wrapfigure}

\paragraph{Multimodal task spaces are broader.}
From an information-theoretic perspective, perplexity measures how well a model trained on one dataset predicts another's distribution~\citep{rosenfeld2002two}. 
We fine-tune a Llama-3.1 8B model separately on the task summaries from each modality (multimodal vs. text-only), and
then evaluate per-sample perplexity on the other modality's task summaries. 
Because all summaries are short (\textless50 words) with low length variance, sequence-length effects are controlled~\citep{xia2023training}. Lower perplexity implies that the training dataset better ``covers'' the task space of the other~\citep{zhao2024wildchat}. Cross-dataset asymmetries in perplexity thus indicate diversity differences: if dataset A models dataset B well but not vice versa, then B contains a broader or more heterogeneous task space. 
As shown in Figure~\ref{fig:perplexity}, models trained on multimodal data achieve lower average perplexity when evaluated on text-only summaries than vice versa (65.55 vs.\ 76.61; Mann–Whitney $U$, $p<0.001$). 

We corroborate this geometrically using out-of-distribution (OOD) detection in embedding space with KNN distance~\citep{sun2022out, zhou2021contrastive} and Isolation Forest~\citep{liu2008isolation, bejan2023ad, cao2025tad}, each trained on one modality and evaluated on the other. As Table~\ref{tab:ood_combined} shows, ooth methods exhibit a clear directional asymmetry: detectors trained on text-only embeddings classify a substantially larger fraction of multimodal tasks as OOD (76\% under KNN and up to $\sim$30\% under Isolation Forest) than the reverse (43\% under KNN and $<$1\% under Isolation Forest). The asymmetry persists after removing modality-specific vocabulary (e.g., ``image'' and ``chart''), indicating that the broader task space is not merely a lexical artifact. Details of the OOD methods, hyperparameter settings, and robustness analyses are provided in Appendix~\ref{appendix:results}.

\begin{table}[t]
\centering
\fontsize{8}{10}\selectfont
\setlength{\tabcolsep}{4pt}
\caption{Out-of-distribution (OOD) detection results on task-summary embeddings. KNN reports the fraction of test instances identified as OOD using nearest-neighbor distance, while Isolation Forest reports the OOD fraction for three contamination settings (with $n_{\text{est}}=200$).}
\label{tab:ood_combined}
\begin{tabular}{lcccc}
\toprule
\textbf{Train $\rightarrow$ Test} & \textbf{KNN (\%)} & \multicolumn{3}{c}{\textbf{Isolation Forest ($n_{\text{est}}{=}200$)}} \\
\cmidrule(lr){3-5}
& & 0.001 & 0.005 & 0.010 \\
\midrule
Text $\rightarrow$ Multimodal & 76.30 & 7.58 & 19.90 & 29.63 \\
Multimodal $\rightarrow$ Text & 42.77 & 0.02 & 0.18 & 0.37 \\
\bottomrule
\end{tabular}
\end{table}

\begin{figure}[t]
    \centering
    \makebox[\textwidth][c]{%
    \includegraphics[width=\linewidth]{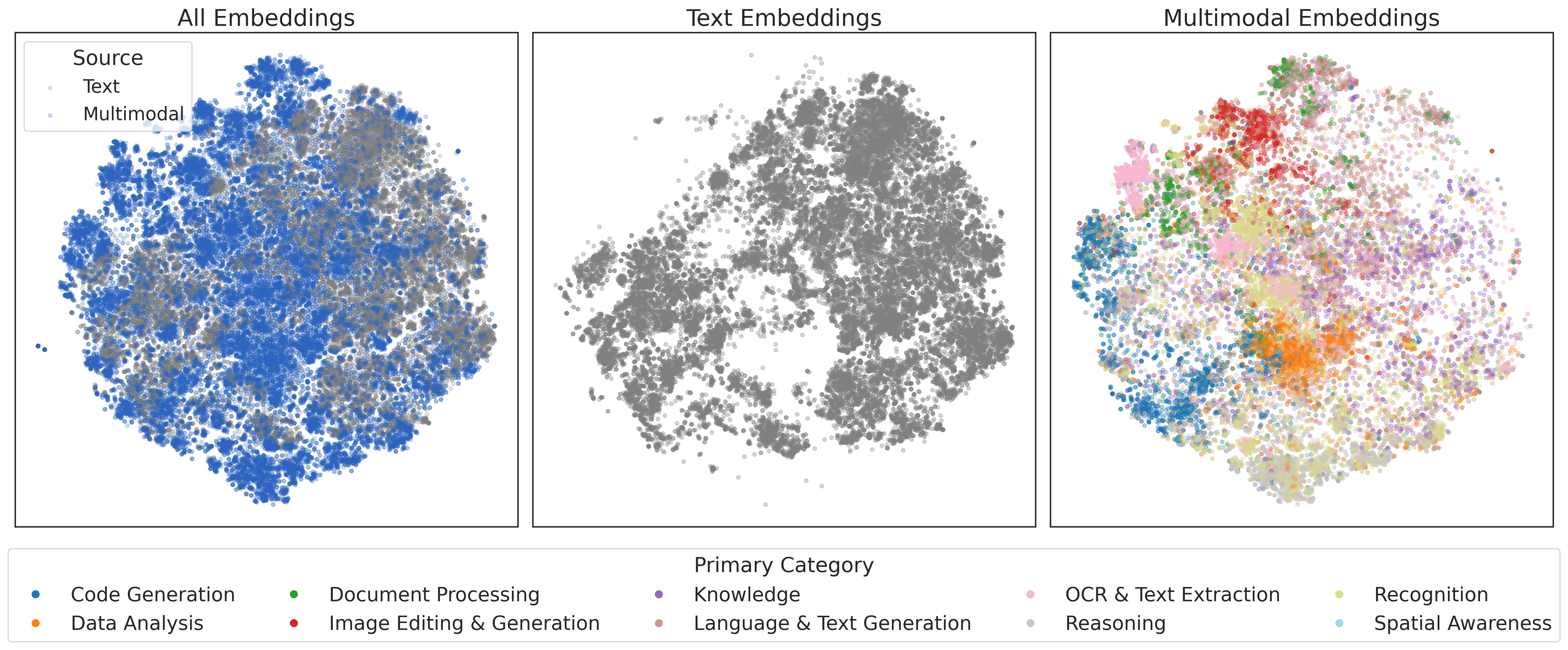}
    }
    \caption{t-SNE of task-summary embeddings. Multimodal tasks (right) occupy a broader, more dispersed region than text-only tasks (middle), forming distinct clusters (OCR, Image Editing \& Generation, Data Analysis, Recognition).}
    \label{fig:tsne_copilot}
\end{figure}

\paragraph{Task-space expansion arises from cross-modal grounding.}
The t-SNE projection (Figure~\ref{fig:tsne_copilot}) corroborates both the perplexity and OOD analyses: although multimodal and text-only task spaces overlap substantially, multimodal embeddings extend beyond the text-only region into distinct clusters. A per-capability analysis (Appendix~\ref{appendix:results}) shows that \emph{Code Generation}, \emph{Document Processing}, and \emph{OCR \& Text Extraction} exhibit the highest OOD ratios, indicating that these capabilities contribute most to the expansion. The most frequent OOD capability combinations (e.g., \emph{Recognition + Reasoning} and \emph{Recognition + Code Generation}; Appendix~\ref{appendix:combination_examples}, Table~\ref{tab:ood_combinations_long}) further suggest that the expansion arises less from visual input alone than from how users compose visual grounding with downstream reasoning and generation.
These findings motivate examining whether existing multimodal benchmarks adequately cover the capability combinations observed in real-world use.

\paragraph{A substantial share of tasks genuinely require visual input.  }
Perplexity and OOD characterize differences between task distributions but do not establish whether individual tasks actually depend on visual input. We therefore conduct a human audit of 200 multimodal sessions, asking two annotators whether each task could reasonably be completed from text alone. On average, \textbf{30\%} of sessions ($60/200$) are judged \emph{uniquely multimodal}, providing direct evidence that a substantial share of real-world tasks require visual grounding rather than merely containing images. The annotation protocol is described in Appendix~\ref{appendix:human_validation}.

\subsection{Multimodal Evaluation Benchmark Analysis}
\label{sec:benchmark}

\definecolor{covStrong}{HTML}{CDE9CE}
\definecolor{covLimited}{HTML}{FCE3B4}
\definecolor{covIncid}{HTML}{F6C4C5}
\newcommand{\covbadge}[2]{\cellcolor{#1}\textbf{#2}}

\begin{table}[t]
\centering
\fontsize{9pt}{9.5pt}\selectfont
\setlength{\tabcolsep}{2pt}
\renewcommand{\arraystretch}{1.25}

\makebox[\textwidth][c]{%
\resizebox{0.99\textwidth}{!}{%
\begin{tabular}{
@{}
>{\raggedright\arraybackslash}p{2.25cm}
>{\raggedright\arraybackslash}p{1.50cm}
>{\raggedright\arraybackslash}p{1.30cm}
>{\raggedright\arraybackslash}p{1.30cm}
>{\raggedright\arraybackslash}p{0.75cm}
>{\raggedright\arraybackslash}p{1.50cm}
>{\raggedright\arraybackslash}p{8.00cm}
@{}
}
\toprule
\textbf{Capability combination}
& \textbf{Real-world freq.}
& \textbf{\# Bench.}
& \textbf{\# Tasks}
& \textbf{S/D ratio}
& \textbf{Coverage}
& \textbf{Representative benchmark tasks} \\
\midrule

\rowcolor{covIncid!30}
Knowledge + Language Generation
& 3{,}563 (8.4\%)
& 13 (5.14\%)
& 16 (22.86\%)
& 0.22
& \covbadge{covIncid}{Incidental}
& MME-Finance~\citep{gan2025mme}: Financial trend explanation;
LLaVA-Bench~\citep{liu2023llava}: Science question explanation;
ScienceQA~\citep{lu2022scienceqa}: Science background generation;
Gut-VLM~\citep{khanal2025hallucination}: Endoscopy report generation \\

Recognition + Reasoning
& 3{,}437 (8.1\%)
& 108 (42.69\%)
& 218 (36.76\%)
& 3.12
& \covbadge{covStrong}{Strong}
& NativeEmbodied~\citep{peng2026foundational}: Object manipulation;
MC-Bench~\citep{xu2025mc}: Multi-image visual grounding;
SciVQA 2025~\citep{borisova2025scivqa}: Scientific figure reasoning;
VCog-Bench~\citep{cao2024visual}: Visual matrix reasoning;
VL-ICLBench~\citep{zong2025vl}: Visual object counting \\

Knowledge + Recognition
& 3{,}330 (7.8\%)
& 54 (21.34\%)
& 85 (29.31\%)
& 1.26
& \covbadge{covStrong}{Strong}
& IS-Bench~\citep{lu2026bench}: Household safety risk recognition;
MIR~\citep{du2025easy}: Visual satire interpretation;
CVQA~\citep{romero2024cvqa}: Culturally grounded visual question answering;
MM-SAP~\citep{wang2024mmsap}: Knowledge-based visual recognition;
GAVIE~\citep{liu2024mitigating}: Visual commonsense reasoning \\

Recognition + Language Generation
& 2{,}291 (5.4\%)
& 52 (20.55\%)
& 89 (26.81\%)
& 1.91
& \covbadge{covStrong}{Strong}
& Centurio~\citep{geigle2025centurio}: Multilingual image captioning;
MDVP-Bench~\citep{lin2025draw}: Point-guided image captioning;
MIA-Bench~\citep{qian2025miabench}: Constrained image description;
Asclepius~\citep{liu2024asclepius}: Medical report generation;
ChartMind~\citep{wei2025chartmind}: Chart classification and explanation \\

Knowledge + Reasoning
& 2{,}132 (5.0\%)
& 58 (22.92\%)
& 83 (23.58\%)
& 1.92
& \covbadge{covStrong}{Strong}
& MMIE~\citep{xia2025mmie}: Knowledge-intensive multi-step reasoning;
GMAI-MMBench~\citep{chen2024gmai}: Medical image question answering;
MMBench~\citep{liu2024mmbench}: Visual attribute reasoning;
MMStar~\citep{chen2024mmstar}: Scientific and technical reasoning;
MathVista~\citep{lu2024mathvista}: Visual math and science reasoning \\

\rowcolor{covIncid!30}
Code Generation + Recognition
& 1{,}643 (3.9\%)
& 3 (1.19\%)
& 3 (23.08\%)
& 0.09
& \covbadge{covIncid}{Incidental}
& ChartX~\citep{xia2024chartx}: Chart reconstruction from images;
Plot2Code~\citep{wu2024plot2code}: Plot-to-code generation \\

\rowcolor{covLimited!35}
Data Analysis + Recognition
& 1{,}150 (2.7\%)
& 9 (3.56\%)
& 9 (20.93\%)
& 0.39
& \covbadge{covLimited}{Limited}
& DynamicVL~\citep{xuan2025dynamicvl}: Visual change estimation;
MULTI~\citep{zhu2025multi}: Diagram and chart question answering;
AutoPaperBench~\citep{kim2025autopaperbench}: Scientific figure comprehension;
MMAU~\citep{sakshi2025mmau}: Scientific figure question answering;
CharXiv~\citep{wang2024charxiv}: Chart information extraction \\

OCR + Recognition
& 985 (2.3\%)
& 30 (11.86\%)
& 40 (29.63\%)
& 2.00
& \covbadge{covStrong}{Strong}
& MMDocBench~\citep{zhu2026mmdocbench}: Document tampering detection;
MEGA-Bench~\citep{chen2025mega}: Visual information extraction;
MMC-Benchmark~\citep{liu2024mmc}: Chart domain classification;
VisualWebArena~\citep{koh2024visualwebarena}: Web information retrieval;
Web2Code~\citep{yun2024web2code}: Webpage screenshot understanding \\

\rowcolor{covLimited!35}
OCR + Language Generation
& 906 (2.1\%)
& 7 (2.77\%)
& 8 (20.51\%)
& 0.43
& \covbadge{covLimited}{Limited}
& AncientDoc~\citep{yu2026ancientdoc}: Classical-to-modern Chinese translation;
TextCaps~\citep{sidorov2020textcaps}: Scene-text-aware image captioning;
WebRRSBench~\citep{liu2026benchmarking}: Web form filling;
LaVy-Bench~\citep{tran2024lavy}: Vietnamese visual question answering;
VisualWebBench~\citep{liu2024visualwebbench}: Webpage summary generation \\

\rowcolor{covIncid!30}
Code Generation + Reasoning
& 859 (2.0\%)
& 2 (0.79\%)
& 2 (10.53\%)
& 0.11
& \covbadge{covIncid}{Incidental}
& MME~\citep{fu2026mme}: Visual code reasoning;
VisualAgentBench~\citep{liu2024visualagentbench}: Screenshot-guided CSS editing \\

\bottomrule
\end{tabular}%
}%
}

\caption{\small
Coverage of the ten most frequent real-world capability combinations.
Parentheses report the proportion of real-world sessions containing each combination for \emph{Real-world freq.}, the proportion of benchmark papers containing at least one relevant task for \emph{\# Bench.}, and the proportion of relevant tasks within those benchmarks for \emph{\# Tasks}.
The S/D ratio compares each combination's share of all eligible benchmark tasks with its share of real-world sessions.
Coverage is \greenhl{Strong} for $\mathrm{S/D} \geq 1$,
\yellowhl{Limited} for $0.3 \leq \mathrm{S/D} < 1$, and
\redhl{Incidental} for $\mathrm{S/D} < 0.3$.
}
\label{tab:coverage}
\end{table}

Our third research objective asks whether existing multimodal benchmarks evaluate the capabilities and capability combinations that users actually invoke in the wild. We map the taxonomy from Section~\ref{sec:taxonomy} onto a large corpus of published multimodal large language model (MLLM) benchmarks and compare benchmark coverage with the real-world demand characterized in Sections~\ref{sec:taxonomy} and~\ref{sec:distribution}.

\paragraph{Benchmark corpus and capability labeling.} 
We draw candidate benchmarks from the survey of \citet{li2024survey}, whose companion
repository\footnote{\url{github.com/swordlidev/Evaluation-Multimodal-LLMs-Survey}, updated May 2026.} provides a comprehensive catalog of MLLM evaluation benchmarks. We retrieve 253 unique benchmark papers and use \texttt{GPT-5.4} to extract 1{,}173 evaluation tasks, as a single benchmark may contain multiple tasks. Two LLM annotators (\texttt{GPT-5.4} and \texttt{Gemini-3.1-Pro}) independently determine whether each task involves static-image input and, if so, which of the ten capabilities it requires. Disagreements are adjudicated by a third model (\texttt{GPT-5.6-sol}), and the resulting labels are validated through a blinded human audit. After excluding non-image tasks and 17 umbrella tasks spanning five or more capabilities, 866 tasks remain. Appendix~\ref{appendix:human_validation} provides the full annotation and validation procedure.

\paragraph{Coverage metric.} 
We quantify coverage using a supply-to-demand (S/D) ratio, comparing each capability pair's share of benchmark tasks with its share of real-world sessions:
\begin{equation}
\mathrm{S/D}_i =
\frac{n^{\mathrm{bench}}_i / N_{\mathrm{bench}}}
     {n^{\mathrm{real}}_i / N_{\mathrm{real}}},
\end{equation}
where $n^{\mathrm{bench}}_i$ and $n^{\mathrm{real}}_i$ denote the number of usage-aligned benchmark tasks and real-world sessions containing capability pair $i$, respectively, and $N_{\mathrm{bench}}$ and $N_{\mathrm{real}}$ denote the corresponding corpus totals. We classify coverage as \textsc{Strong} ($\mathrm{S/D}_i\!\geq\!1$), \textsc{Limited} ($0.3\!\leq\!\mathrm{S/D}_i\!<\!1$), or \textsc{Incidental} ($\mathrm{S/D}_i\!<\!0.3$). Table~\ref{tab:coverage} reports the results.

\paragraph{Benchmarks favor image interpretation over artifact construction.}
Coverage is strongest exactly where an image maps to a well-defined answer. \textit{Recognition + Reasoning} ($S/D\!=\!3.12$), \textit{Knowledge + Reasoning} ($1.92$), \textit{Recognition + Language Generation} ($1.91$), \textit{Knowledge + Recognition} ($1.26$), and \textit{OCR + Recognition} ($2.00$) are all \textsc{Strong}: they mirror users' requests to identify what an image shows and reason toward a fixed answer. Coverage is smaller when the image is instead the starting point for a new, usable output such as a document, a program, or a decision, a gap that takes two forms.

\paragraph{Generative writing tasks are evaluated only as short-form answers.}
\textit{Knowledge + Language \& Text Generation} is the single most frequent real-world combination (8.4\% of sessions) yet only \textsc{Incidental} ($S/D\!=\!0.22$). Its matching benchmark tasks stop at a short explanation or answer, whereas real users read a document, screenshot, or note and produce a full piece of structured writing, such as a research summary, a policy, or an explanatory report grounded in the relevant subject knowledge. \textit{OCR + Language \& Text Generation} shows the same gap (\textsc{Limited}, $S/D\!=\!0.43$): its real usage also require reading the extracted text and then composing a fluent, constraint-respecting document. In each case, benchmarks test the reading step but rarely the sustained writing that follows.

\paragraph{Code and data generation from images remains largely untested.}
Code and data workflows are the scarcest. \textit{Code Generation + Recognition} ($S/D\!=\!0.09$) and \textit{Code Generation + Reasoning} ($S/D\!=\!0.11$) together match only five tasks, all of which read or lightly edit existing code, such as reconstructing a plot or editing CSS to match a screenshot, whereas users build full artifacts from visual input by generating interfaces from mockups and implementing algorithms from photographed problems. \textit{Data Analysis + Recognition} (\textsc{Limited}, $S/D\!=\!0.39$) likewise stops at reading charts, while real requests go on to compute, compare, and recommend actions from them. 

\paragraph{Implications for benchmark design.}
Benchmarks are strongest when the target output is a label, value, caption, or short explanation, but real multimodal use routinely continues past these endpoints, to a document rewritten for an audience, a mockup turned into working software, or a chart turned into a decision. Closing the gap calls for benchmarks built around complete visual-to-artifact workflows, scored on the fidelity, executability, and decision relevance of the produced artifact rather than on a single correct answer. Such task families would complement, not replace, current perception and reasoning benchmarks.

\section{Discussion and Conclusion}

This study provides one of the first large-scale empirical characterizations of how people use multimodal LLMs in the wild, focusing on sessions where users upload images. Drawing on conversations from Microsoft Copilot and ChatGPT, we introduce a hierarchical taxonomy of ten multimodal capabilities spanning perception, cognition, and generation, and show that image-upload tasks are predominantly compositional: most sessions couple perceptual entry points with downstream reasoning and generation. We further find that multimodal interactions occupy a broader task space than text-only ones, driven by cross-modal grounding rather than the presence of images alone, and that current benchmarks concentrate on perception and reasoning while the compositional generation workflows that dominate real use remain undertested.

\paragraph{A gap between how multimodal LLMs are evaluated and used.}
Existing benchmarks are largely built top-down from expert taxonomies and evaluated under controlled settings, concentrating on tasks where an image maps to a well-defined answer~\citep{yu2024mm, hao2025can, ramachandran2025does}. Real usage rarely stops there: \textit{Recognition}, \textit{OCR \& Text Extraction}, and \textit{Knowledge} typically serve as entry points into multi-step workflows that terminate in generation. This gap is uneven across capabilities. Combinations that read an image and reason toward a fixed answer are well covered, whereas the writing-, code-, and data-generation workflows common in real-world use, led by the single most frequent combination \textit{Knowledge + Language \& Text Generation}, are evaluated only incidentally. Closing this gap calls for benchmarks organized around complete visual-to-artifact workflows, built from observed task chains (e.g., extract $\rightarrow$ interpret $\rightarrow$ generate) and grounded in usage frequency rather than research convention, and scored on the fidelity, executability, and decision relevance of the produced output rather than on a single correct answer. A complementary direction is to study how upstream perception errors (e.g., OCR) propagate to downstream reasoning and generation.

\paragraph{The cognitive character of multimodal use and its implications for interface design.}
Two findings suggest that image uploads reshape the cognitive profile of a task rather than simply re-expressing it in another modality. First, users employ images to reduce effort: frequent uploads of structured textual content such as ``code snippet,'' ``HTML/CSS,'' and ``Excel/table'' (Table~\ref{tab:tfidf_keywords}) indicate that users share screenshots rather than retype formatted text, offloading the burden of reproducing layout and context that is cumbersome to transcribe. This is consistent with \textit{cognitive offloading}~\citep{risko2016cognitive} and with prior work identifying screenshots as lightweight tools for memory and communication~\citep{cramer2023s}. Second, the tasks themselves are more demanding: classifying sessions under Anderson and Krathwohl's revised taxonomy of cognitive processes~\citep{wilson2016anderson}, we find multimodal sessions skew toward higher-order activities (e.g., Apply, Analyze, Create) relative to text-only sessions ($\chi^2$ test, $p<0.001$; Appendix~\ref{appendix:results}). Image uploads are thus associated with more complex, generative objectives, expanding the functional scope of what users attempt. These patterns point to a shared interface implication: systems should treat such uploads as first-class structured input, recovering the underlying code, table, or layout rather than a flat image, and accept mixed image-and-text prompts, so that they lower the friction of moving formatted content into a conversation and support the more generative workflows that images tend to accompany.

\paragraph{Conclusion.}
By grounding capability taxonomies and benchmark design in observed user behavior, this work offers an empirical basis for evaluations and interfaces that reflect real-world needs. Our findings suggest that the next generation of multimodal benchmarks should be organized less around tasks that map an image to a fixed answer and more around the compositional, generation-terminating workflows that characterize usage in the wild, and that interfaces should be designed to support the offloading and generative objectives these workflows entail.


\paragraph{Limitations.}
Our study has several limitations. First, due to privacy constraints, the analysis is conducted in an \emph{eyes-off} setting: we rely on privacy-preserving task and image summaries, which may omit fine-grained visual details. Our eyes-on validation indicates that these summaries preserve the capability signal with limited loss. Second, our findings may not fully generalize to future multimodal systems, as advances in model capabilities and interfaces could change how people interact with LLMs or give rise to new task categories. To mitigate this concern, we validate our taxonomy and findings on two independent datasets spanning different AI assistants and collection periods, providing evidence that the taxonomy is not specific to the Copilot sample; longitudinal studies will nonetheless be important for tracking how usage evolves. Finally, privacy constraints prevent releasing the original conversations, making the development of methods for sharing realistic yet privacy-preserving datasets an important direction for future work.


\section*{Ethical Considerations}
All data analyzed in this study were collected and processed in accordance with established privacy, security, and ethical standards. The primary dataset consists of de-identified Copilot conversations analyzed within secure, access-controlled research environments, and the validation dataset was obtained from an independent provider under appropriate data protection safeguards and informed user consent.

All conversations were de-identified prior to analysis, with automated processing removing personally identifiable information such as names, contact details, and other sensitive attributes. To further protect privacy, the analysis was conducted under an \emph{eyes-off} setting: conversations were transformed into concise task and image summaries without direct access to raw images or reproduction of original user text, and taxonomy induction, annotation, and all subsequent analyses operated exclusively on these summaries. We made no attempt to re-identify users or infer sensitive personal attributes, the study did not involve interaction with human participants, and all results are reported only in aggregate.

\bibliography{reference}
\bibliographystyle{colm2026_conference}

\newpage
\input{main_appendix.tex}

\end{document}

%% file: main_appendix.tex
\appendix

\section{Extended Results}
\label{appendix:results}

\subsection{Additional Information about the Copilot Dataset}
\label{appendix:results_copilot}

    

\begin{figure}[h]
    \centering

    \begin{subfigure}{0.6\linewidth}
        \centering
        \includegraphics[width=\linewidth]{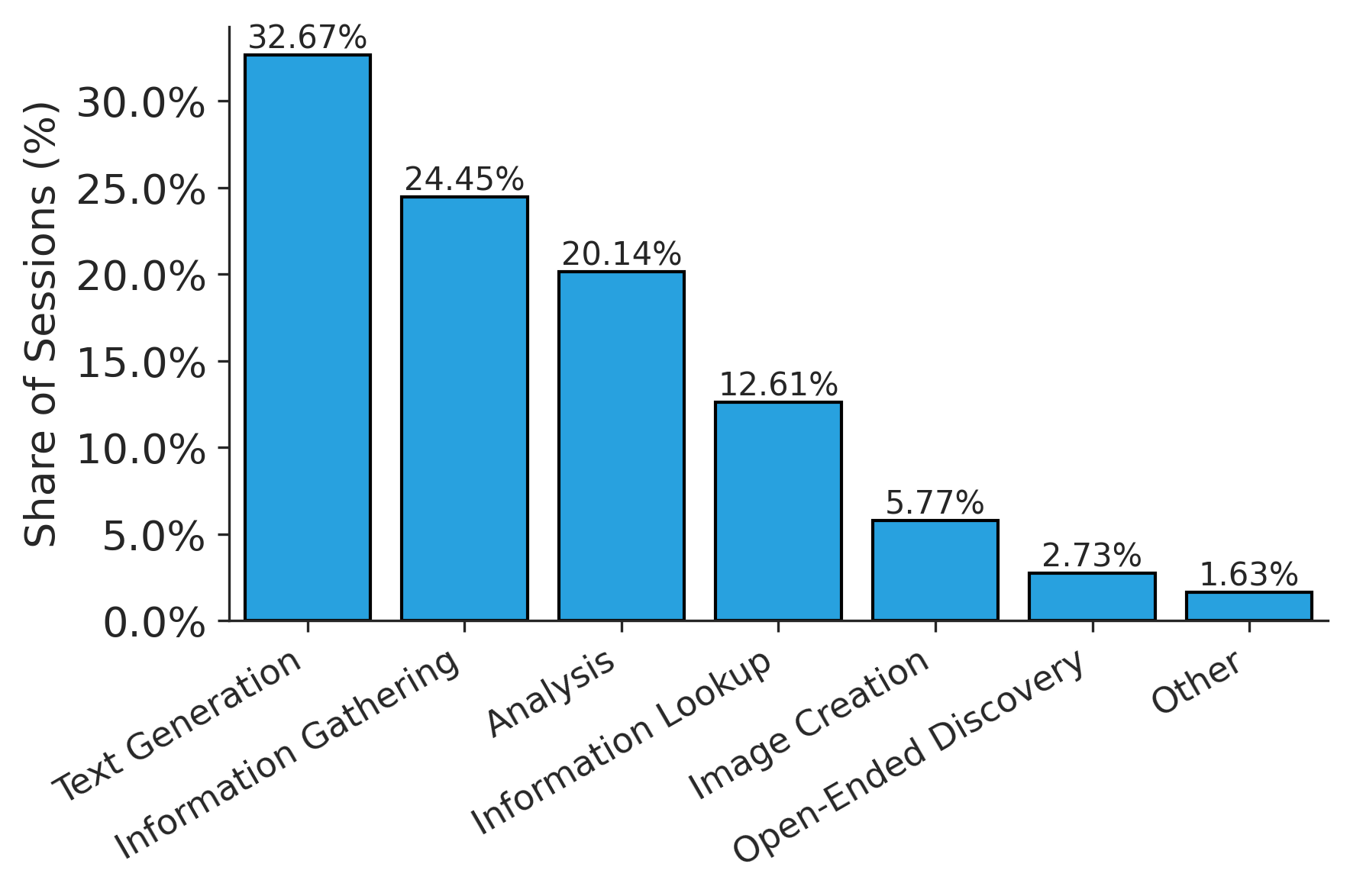}
    \end{subfigure}

    \vspace{0.5em}

    \begin{subfigure}{0.6\linewidth}
        \centering
        \includegraphics[width=\linewidth]{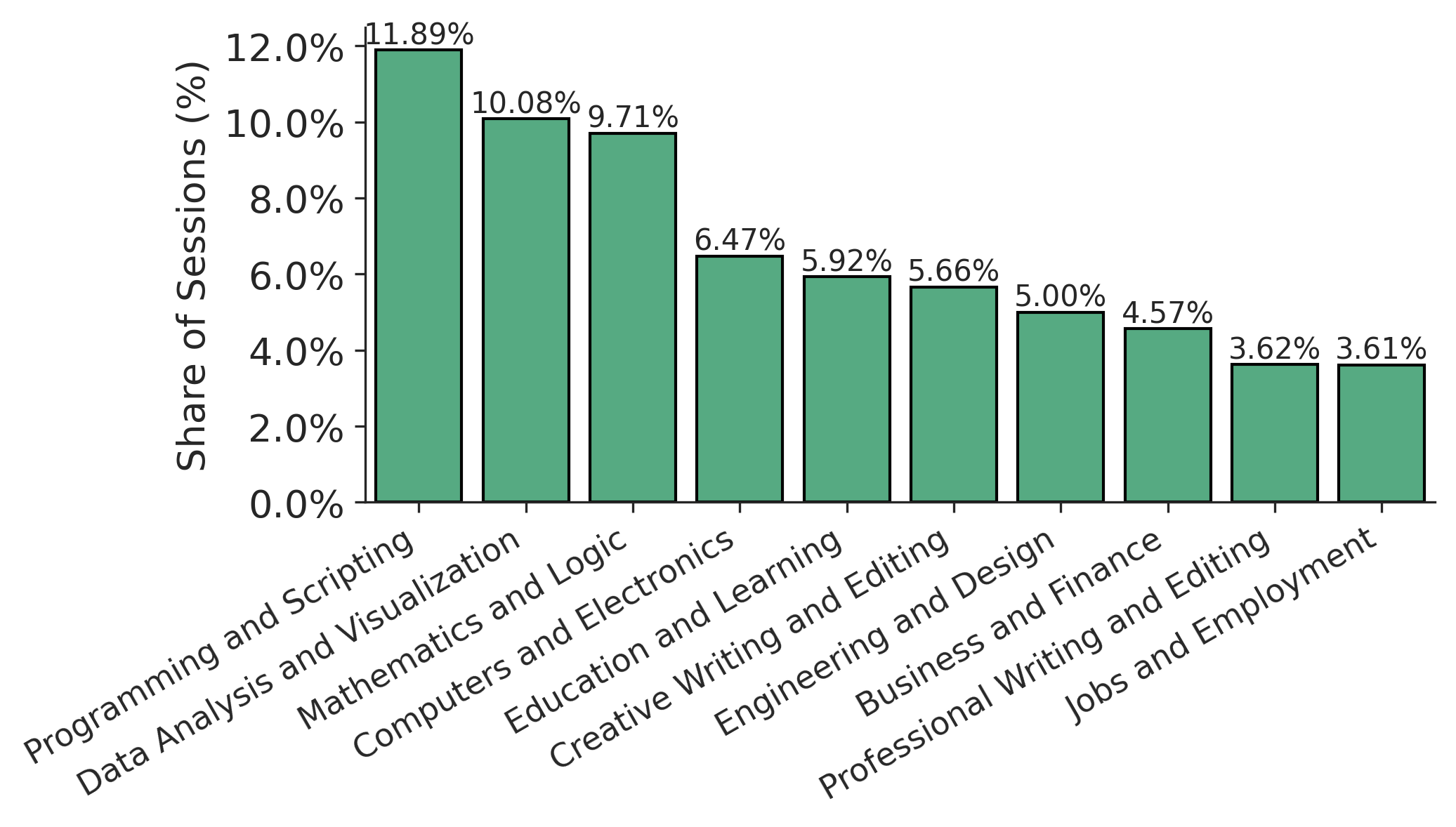}
    \end{subfigure}

    \caption{Frequency distribution of user intents (top) and distribution of sessions across the top 10 primary domains (bottom).}
    \label{fig:intents_domains}
\end{figure}

To analyze user intents and conversation domains, we apply an LLM-based taxonomy generator and classifier developed by \citet{wan2024tnt}. The distributions of user intents and primary domains in the multimodal subset of the Copilot dataset are reported in Figure~\ref{fig:intents_domains}. We note that the dataset reflects an early adoption phase following the introduction of visual capabilities in Copilot, which may influence observed usage patterns.

\paragraph{Out-of-distribution Detection.}

\begin{table}[t]
\centering
\caption{Out-of-distribution (OOD) detection results in embedding space.}
\label{tab:ood_combined_full}
\fontsize{8}{9}\selectfont

\begin{subtable}{\linewidth}
\centering
\begin{tabular}{lc}
\toprule
\textbf{Train $\rightarrow$ Test} & \textbf{KNN (\%)} \\
\midrule
Text $\rightarrow$ Multimodal & 76.30 \\
Multimodal $\rightarrow$ Text & 42.77 \\
\bottomrule
\end{tabular}
\caption{KNN OOD fraction.}
\label{tab:ood_knn}
\end{subtable}

\vspace{0.8em}

\begin{subtable}{\linewidth}
\centering
\begin{tabular}{lcccc}
\toprule
\textbf{Train $\rightarrow$ Test} & \textbf{$n_{\text{est}}$} & \multicolumn{3}{c}{\textbf{Contamination}} \\
\cmidrule(lr){3-5}
 & & \textbf{0.001} & \textbf{0.005} & \textbf{0.010} \\
\midrule
\multirow{3}{*}{Text $\rightarrow$ Multimodal}
& 50  & 3.97 & 11.43 & 17.25 \\
& 100 & 4.94 & 15.30 & 22.43 \\
& 200 & 7.58 & 19.90 & 29.63 \\
\midrule
\multirow{3}{*}{Multimodal $\rightarrow$ Text}
& 50  & 0.03 & 0.14 & 0.27 \\
& 100 & 0.02 & 0.12 & 0.26 \\
& 200 & 0.02 & 0.18 & 0.37 \\
\bottomrule
\end{tabular}
\caption{Isolation Forest OOD fraction.}
\label{tab:ood_iso}
\end{subtable}

\end{table}

While perplexity measures how well one task distribution predicts another, it does not capture the geometric relationship between task spaces. We therefore perform out-of-distribution (OOD) detection on task-summary embeddings generated by \texttt{text-embedding-3-large}. Because these embeddings are high-dimensional and non-Gaussian, we adopt two widely used nonparametric detectors: $k$-nearest neighbor (KNN) distance~\citep{sun2022out, zhou2021contrastive} and Isolation Forest~\citep{liu2008isolation, bejan2023ad, cao2025tad}. Each detector is trained on one modality (treated as the in-distribution) and evaluated on the other. For Isolation Forest, we grid-search $n_{\text{estimators}}\in\{50,100,200\}$ and contamination rates $\{0.001,0.005,0.01\}$.

The complete OOD results are reported in Table~\ref{tab:ood_combined_full}. Consistent with the findings reported in the main paper, both detectors exhibit strong directional asymmetry, indicating that multimodal task representations cover a broader semantic space than text-only interactions.

\paragraph{Per-capability OOD Analysis.}

To identify which capabilities contribute most to the expansion of multimodal task space, we compute OOD ratios separately for each capability using Isolation Forest ($n_{\text{estimators}}=100$, contamination$=0.005$). OOD counts are aggregated over both primary and secondary capability labels.

Table~\ref{tab:ood_per_category} shows substantial variation across capabilities. The highest OOD ratios occur for \textit{Code Generation}, \textit{Document Processing}, and \textit{OCR \& Text Extraction}, indicating that these capabilities contribute most strongly to the semantic expansion beyond text-only interactions.

\begin{table*}[t]
\centering
\small
\caption{Per-category OOD rates (Isolation Forest, $n_{\text{estimators}}=100$, contamination=0.005). Counts are computed over both primary and secondary labels.}
\label{tab:ood_per_category}
\begin{tabular}{lccc}
\toprule
\textbf{Capability} & \textbf{OOD Count} & \textbf{Total Count} & \textbf{OOD Ratio} \\
\midrule
Code Generation & 998 & 4,253 & 0.235 \\
Document Processing & 492 & 2,230 & 0.221 \\
OCR \& Text Extraction & 749 & 3,804 & 0.197 \\
Reasoning & 1,606 & 8,676 & 0.185 \\
Image Editing \& Generation & 174 & 973 & 0.179 \\
Data Analysis & 623 & 4,074 & 0.153 \\
Recognition & 1,643 & 13,433 & 0.122 \\
Language \& Text Generation & 1,194 & 9,890 & 0.121 \\
Knowledge & 1,358 & 11,710 & 0.116 \\
Spatial Awareness & 13 & 139 & 0.094 \\
\bottomrule
\end{tabular}
\end{table*}

\paragraph{Robustness to Modality Vocabulary.}

A potential concern is that the observed asymmetry arises from modality-specific vocabulary rather than differences in task semantics. To evaluate this possibility, we remove explicit modality tokens (e.g., ``image,'' ``picture,'' ``photo,'' ``screenshot,'' ``visual,'' ``chart,'' ``table,'' and ``diagram'') from multimodal task summaries, recompute embeddings, and repeat the analyses.

The directional asymmetry persists across both detectors. Averaged over Isolation Forest settings, Text$\rightarrow$Multimodal OOD rates remain substantially higher than the reverse direction (15.24\% vs.\ 0.16\%). KNN exhibits the same pattern (77.09\% vs.\ 41.46\%). These results indicate that the broader multimodal task space is not merely a lexical artifact.

\paragraph{OOD Capability Combinations.}

The expansion of multimodal task space is not explained solely by structured visual inputs. We therefore examine the most frequent capability combinations among OOD sessions. Representative combinations are reported in Table~\ref{tab:ood_combinations_long}. The dominant OOD workflows combine perceptual capabilities with downstream reasoning or generation (e.g., \textit{Recognition + Reasoning}, \textit{Recognition + Code Generation}, and \textit{OCR \& Text Extraction + Language \& Text Generation}), suggesting that semantic expansion arises from cross-modal grounding and capability composition rather than from image modality alone.

\begin{figure}[t]
    \centering
    \includegraphics[width=0.6\columnwidth]{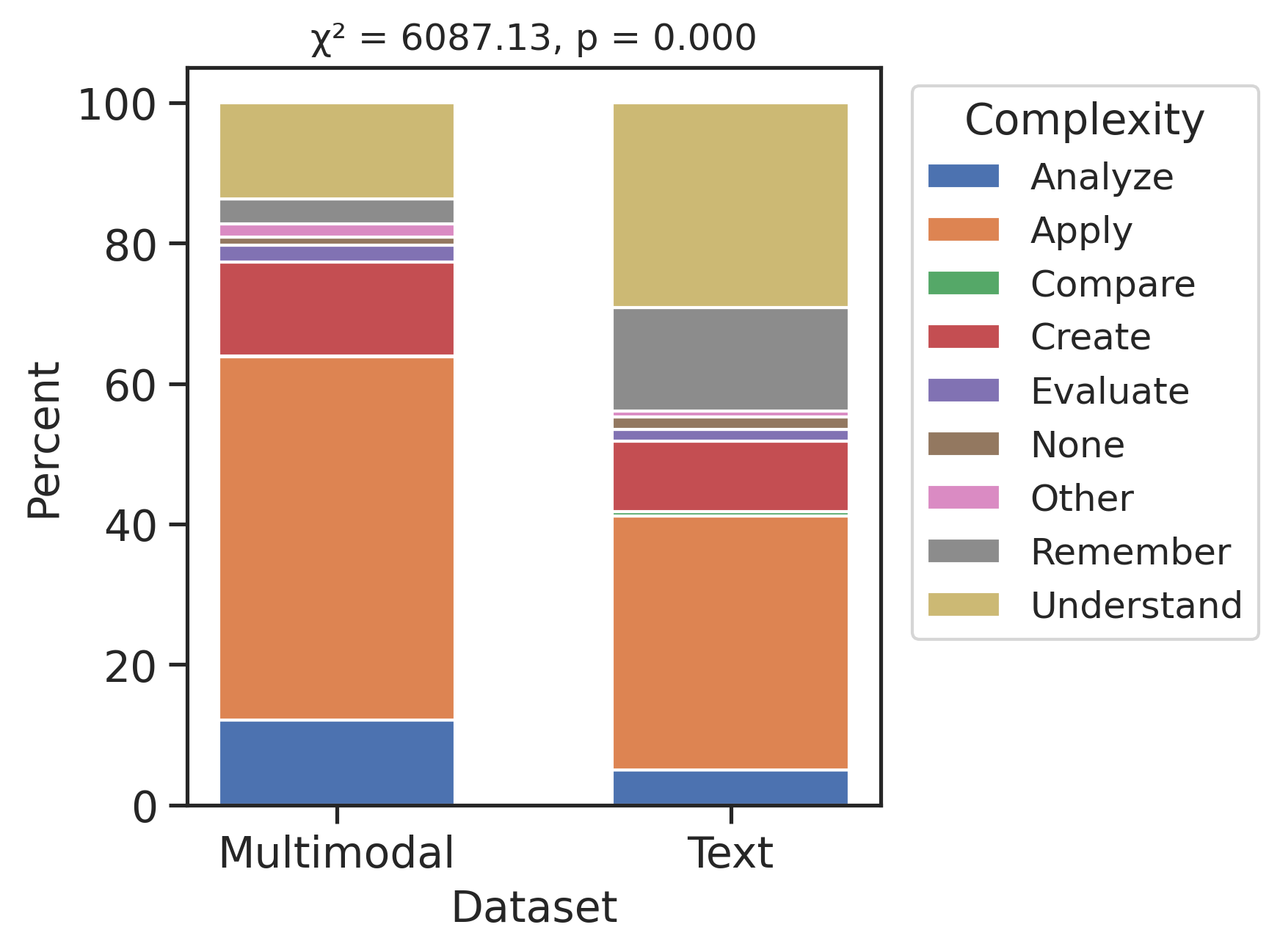}
    \caption{Distribution of task complexity across multimodal and text-only Copilot sessions.}
    \label{fig:complexity_copilot}
\end{figure}

\paragraph{Task Complexity Analysis.}
We measure task complexity using Anderson and Krathwohl's revised taxonomy of cognitive processes~\citep{wilson2016anderson}, following the LLM-based classifier of~\citet{suri2024use}. Each conversation is assigned one of six cognitive levels (\textit{Remember}, \textit{Understand}, \textit{Apply}, \textit{Analyze}, \textit{Evaluate}, and \textit{Create}) according to the effort required for a human to complete the task without AI assistance. Following~\citet{suri2024use}, we further group tasks into \textit{low complexity} (Remember) and \textit{high complexity} (all remaining levels).
A chi-square test reveals a significant difference between multimodal and text-only sessions ($\chi^2(8)=6087.13$, $p<.001$). As shown in Figure~\ref{fig:complexity_copilot}, multimodal sessions contain substantially more higher-order cognitive activities, particularly \textit{Apply} (51.79\% vs.\ 36.13\%), \textit{Analyze} (12.05\% vs.\ 5.01\%), and \textit{Create} (13.38\% vs.\ 10.01\%), whereas text-only sessions more frequently involve lower-order tasks such as \textit{Understand} (29.17\%) and \textit{Remember} (14.70\%). These results suggest that multimodal interactions are associated with more complex user objectives.

\subsection{ChatGPT Dataset Validation Results}
\label{appendix:results_chatgpt}

To validate the reliability and generalizability of our taxonomy and findings, we conduct parallel analyses on a second dataset collected from a different AI chatbot: ChatGPT. This dataset spans two months, November 2024 and January 2025, and includes $\sim$732,000 conversation sessions from 7,676 unique users. Among these sessions, 23,413 contain at least one image; the remaining sessions are text-only. Each session is annotated with a variety of metadata fields, including intent labels, domain categories, and an image flag indicating whether the session involved any visual content. 

We note several caveats that limit the dataset's comparability with the Copilot dataset: (i) all interactions were conducted on desktop devices, (ii) users are located exclusively in the United States, and (iii) the dataset does not distinguish between image \textit{uploads} and image \textit{generations}, relying only on a binary flag indicating the presence of any image. Due to these constraints, we treat this dataset as a \emph{validation dataset} rather than a primary source of analysis. We repeat our analyses on this dataset and report the results as follows.

\begin{figure}[t]
    \centering
    \includegraphics[width=0.85\linewidth]{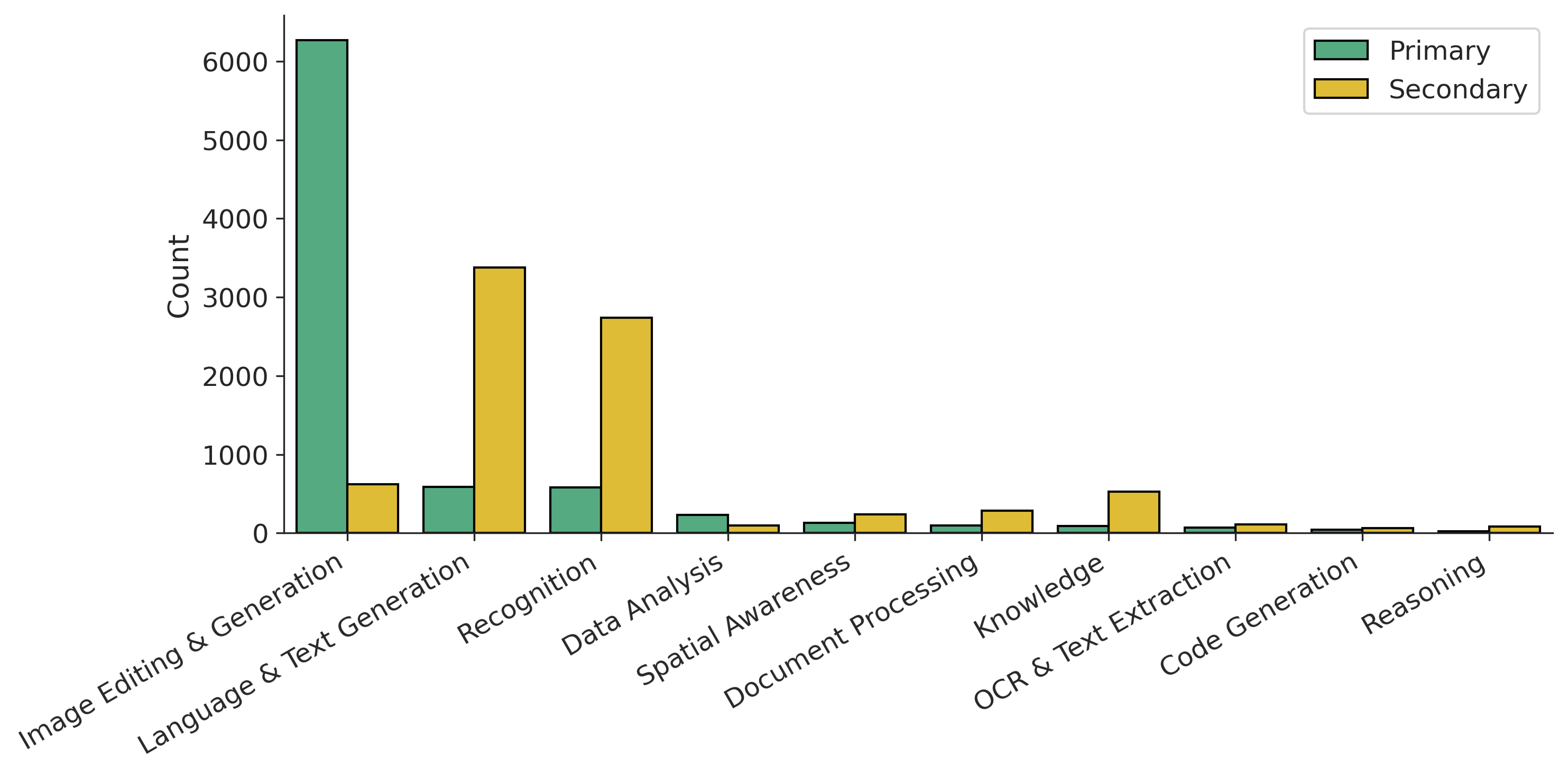}
    \caption{Frequency distributions of primary and secondary capabilities.}
    \label{fig:capability_distribution_chatgpt}
\end{figure}

Applying the Copilot-derived taxonomy to the ChatGPT validation dataset yields a capability distribution with \textit{Image Editing \& Generation}, \textit{Language \& Text Generation}, and \textit{Recognition} as the three most frequent primary capabilities (Figure~\ref{fig:capability_distribution_chatgpt}). \textit{Image Editing \& Generation} dominates the distribution, partly because the ChatGPT dataset does not distinguish between user-uploaded images and model-generated images. As a result, image creation activities are aggregated into the same multimodal category, increasing its relative frequency. Only 62 sessions (0.26\%) are labeled as ``Undefined''. Overall, the distribution supports the cross-platform robustness of the taxonomy. 

\begin{figure*}[t]
    \centering
    \makebox[\textwidth][c]{%
    \includegraphics[width=\linewidth]{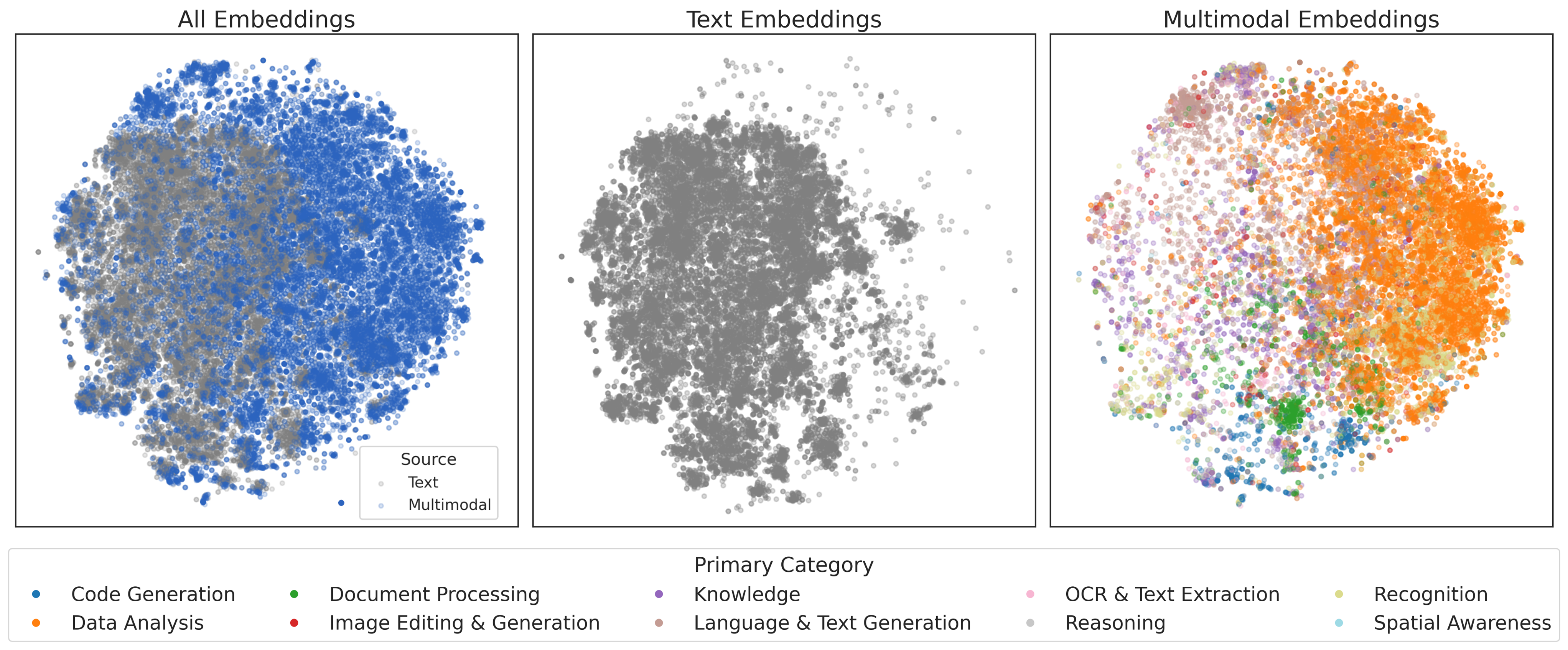}
    }
    \caption{t-SNE visualization of task-summary embeddings across modalities.}
    \label{fig:tsne_chatgpt}
\end{figure*}

We replicate the embedding-space analyses on the ChatGPT validation dataset to test robustness. Table~\ref{tab:ood_combined} shows the same directional asymmetry: models trained on text-only embeddings flag a larger fraction of multimodal samples as outliers than the reverse. This pattern holds across both KNN and Isolation Forest. Figure~\ref{fig:tsne_chatgpt} shows that multimodal embeddings extend beyond text-only regions and form distinct clusters. These results confirm that multimodal tasks occupy a broader and more heterogeneous semantic space.

\begin{table}[t]
\centering
\footnotesize
\caption{Out-of-distribution (OOD) detection results in embedding space.}
\label{tab:ood_combined_chatgpt}

\begin{subtable}{\linewidth}
\centering
\begin{tabular}{lc}
\toprule
\textbf{Train $\rightarrow$ Test} & \textbf{KNN (\%)} \\
\midrule
Text $\rightarrow$ Multimodal & 76.85 \\
Multimodal $\rightarrow$ Text & 54.84 \\
\bottomrule
\end{tabular}
\caption{Fraction of outliers detected using KNN distance.}
\label{tab:ood_knn_chatgpt}
\end{subtable}

\vspace{0.8em}

\begin{subtable}{\linewidth}
\centering
\begin{tabular}{lcccc}
\toprule
\textbf{Train $\rightarrow$ Test} & \textbf{$n_{\text{est}}$} & \multicolumn{3}{c}{\textbf{Contamination}} \\
\cmidrule(lr){3-5}
 & & \textbf{0.001} & \textbf{0.005} & \textbf{0.010} \\
\midrule
\multirow{3}{*}{Text $\rightarrow$ Multimodal}
& 50  & 0.95 & 3.59 & 6.50 \\
& 100 & 1.14 & 4.01 & 7.55 \\
& 200 & 1.27 & 4.87 & 9.05 \\
\midrule
\multirow{3}{*}{Multimodal $\rightarrow$ Text}
& 50  & 0.04 & 0.06 & 0.60 \\
& 100 & 0.06 & 0.07 & 0.68 \\
& 200 & 0.08 & 0.11 & 0.90 \\
\bottomrule
\end{tabular}
\caption{OOD fractions from Isolation Forest with varying numbers of estimators ($n_{\text{est}}$) and contamination levels.}
\label{tab:ood_iso_chatgpt}
\end{subtable}

\end{table}

\newpage
\section{Prompt Templates}
\label{appendix:prompt}

\textbf{Prompt 1: Summarization prompt (Step 1 in Section~\ref{sec:taxonomy}).}

\begin{promptbox}
You are given a transcript of a conversation between a User and an AI assistant that includes user-uploaded images.\\
\vspace{1em}
\textbf{Instructions:}\\
\textbf{1. Extract the User's main objective in no more than 50 words. Format exactly as:}\\
\hspace*{1em}\texttt{The user's main objective is [...].}\\
\hspace*{1em}- \textbf{Output only the text inside the brackets, nothing else.}\\
\hspace*{1em}- \textbf{The phrase inside [ ] must start with a capitalized verb or list a series of action phrases.}\\
\hspace*{1em}- \textbf{Include action details the User specified.}\\
\hspace*{1em}- Example:\\
\hspace*{2em}If the User asks \texttt{"Please crop and center the image, then increase contrast,"} your output is:\\
\hspace*{2em}\texttt{"Crop and center the image, then increase contrast"}\\
\vspace{0.5em}
\textbf{2. Describe how the User leveraged the LLM assistant's multimodal capabilities}, i.e., after uploading the image, what the model needs to do with the image (e.g., recognition, OCR, image generation, etc.) in no more than 50 words.\\
\vspace{1em}
\textbf{Wrap your outputs in XML tags:}\\
\texttt{<summary>\ldots</summary>}\\
\texttt{<capability>\ldots</capability>}\\
\vspace{0.5em}
\textbf{If the transcript is empty or too brief to extract a goal or workflow, output:}\\
\texttt{<summary></summary>}\\
\texttt{<capability></capability>}\\
\vspace{1em}
\textbf{Conversation transcript:}\\
\textcolor{salmonred}{\textbf{\texttt{{\{\{Text\}\}}}}}
\end{promptbox}

\noindent \textbf{Prompt 2: Taxonomy refinement prompt (Step 2 in Section~\ref{sec:taxonomy}).} 

We initialize \texttt{cluster\_table} with categories drawn from prior multimodal benchmarks. The taxonomy is limited by \texttt{max\_num\_clusters = 10}, with each category name constrained by \texttt{cluster\_name\_length = 4} words. Category descriptions are generated in two formats specified by \texttt{cluster\_description\_length}, with a short version capped at 50 words (shown in the main text) and a detailed version capped at 200 words (included in the Appendix). Each iteration also elicits a quality rating, a justification of up to \texttt{explanation\_length = 200} words, and a revision suggestion of up to \texttt{suggestion\_length = 200} words.

\begin{promptbox}
cluster\_table = """ \\
\color{grassgreen}
\texttt{| id | name                | description |} \\
\texttt{|----|---------------------|-------------|} \\
\texttt{| 1  | Recognition         | Recognition refers to the model's ability to visually identify scenes, objects, and attributes such as color, material, and shape, perform counting, and tackle high-level visual recognition tasks. |} \\
\texttt{| 2  | Knowledge           | Knowledge covers the model's capacity to leverage social and visual commonsense, encyclopedic facts, and time-sensitive information (e.g., news) to solve complex, multi-step problems. |} \\
\texttt{| 3  | OCR                 | Optical character recognition (OCR) measures the model's proficiency in reading and understanding text embedded in images and reasoning over that text to complete various tasks. |} \\
\texttt{| 4  | Spatial Awareness   | Spatial awareness captures the model's ability to interpret and reason about precise object positioning, spatial relationships (e.g., "next to," "above," "behind"), and interactions among objects and text regions within an image. |} \\
\texttt{| 5  | Language Generation | Language generation evaluates the model's skill in producing clear, engaging, and informative written responses, especially for prompts requiring extended or nuanced answers. |} \\
\texttt{| 6  | Reasoning           | Reasoning encompasses the model's ability to apply general world knowledge and structured, rule‐based logic to draw inferences, discern complex social, physical, or natural relationships, and perform cause-and-effect analyses over combined visual and textual inputs. |} \\
\color{black}
""" \\
\end{promptbox}

\begin{promptbox}
\textbf{Instructions} \\
- Your input is a set of conversation summaries between a \textbf{User} and an \textbf{AI assistant}, all of which involve \textbf{user-uploaded images}. \\
- Your goal is to classify which multimodal capabilities each conversation leverages, using a provided reference table. \\
- \textbf{Important:} The descriptions in the initial reference table are broad and underspecified. \\
You must: \\
  - \textbf{Review} each category's current description against the example summaries. \\
  - \textbf{Identify} common sub-use cases for each category. \\
  - \textbf{Refine} the descriptions with concrete, specific examples and non-overlapping boundaries. \\
  - \textbf{Iteratively update} the table so it faithfully captures the breadth of real workflows. \\
\vspace{1em}
\textbf{Input user conversation data structure} \\
The input user conversation summary table is a markdown table with each row as a category, with the following columns: \\
- \textbf{id}: conversation index. \\
- \textbf{summary}: summarization of the conversation the user had with the AI agent. \\
\vspace{1em}
\textbf{Input reference table format} \\
The input cluster table is a markdown table with each row as a category, with the following columns: \\
- \textbf{id}: category index. \\
- \textbf{name}: category name. \\
- \textbf{description}: category description used to classify data points. \\
\vspace{1em}
\textbf{Reference cluster table} \\
\textcolor{salmonred}{\textbf{\{\{cluster\_table\}\}}} \\
\vspace{1em}
\textbf{User conversation summary table} \\
\textcolor{salmonred}{\textbf{\{\{Text\}\}}} \\
\vspace{1em}
\textbf{Output Requirements} \\
\textbf{Format}: \\
- Output clusters as a \textbf{markdown table} with each row as a category, with the following columns: \\
  - \texttt{id}: category index starting from 1 in an incremental manner. \\
  - \texttt{name}: category name should be \textbf{within \textcolor{salmonred}{\{\{cluster\_name\_length\}\}} words}. \\
  - \texttt{description}: category description should be \textbf{within \textcolor{salmonred}{\{\{cluster\_description\_length\}\}} words}. \\
- Total number of categories should be \textbf{no more than \textcolor{salmonred}{\{\{max\_num\_clusters\}\}}}. \\
- Output table should be in \textbf{English only}. \\
\vspace{1em}
Example format: \\
\texttt{| id | name | description |} \\
\texttt{|-|-|-|} \\
\texttt{| category id | category name | category description |} \\
\vspace{1em}
\textbf{Quality} \\
- \textbf{No overlap or contradiction} among the categories. \\
- \textbf{Name} is a concise and clear label for the category. \\
- \textbf{Description} differentiates one category from another. \\
- \textbf{Name and description} can \textbf{accurately and consistently} classify new data points \textbf{without ambiguity}. \\
- \textbf{Name and description} are \textbf{consistent with each other}. \\
- Output clusters \textbf{match the data} as closely as possible. \\
- Categories must be \textbf{specific and meaningful}. Avoid vague or catch-all labels. \\
\vspace{1em}
\textbf{Questions} \\
- \textbf{Q1:} Review the given reference table and the input data and provide a \textbf{rating score} of the reference table. The rating score should be an \textbf{integer between 0 and 100}, higher rating score means better quality. \\
- \textbf{Q2:} Explain your rating score in Q1 \textbf{within \textcolor{salmonred}{\{\{explanation\_length\}\}} words}. \\
- \textbf{Q3:} Based on your review, decide if you need to \textbf{edit the reference table} to improve its quality. If yes, suggest potential edits \textbf{within \textcolor{salmonred}{\{\{suggestion\_length\}\}} words}. If no, output "N/A". \\
- \textbf{Q4:} If you decide to edit the reference table, please provide your \textbf{updated reference table}. If you decide not to edit the reference table, please output the original reference table. \\
\vspace{1em}
Wrap your answers as: \\
\texttt{<rating>your answer to Q1 between 0 and 100</rating>} \\
\texttt{<explanation>your answer to Q2 within \textcolor{salmonred}{\{\{explanation\_length\}\}} words</explanation>} \\
\texttt{<suggestion>your answer to Q3 within \textcolor{salmonred}{\{\{suggestion\_length\}\}} words</suggestion>} \\
\texttt{<cluster\_table>your answer to Q4 in markdown table format with no more than \textcolor{salmonred}{\{\{max\_num\_clusters\}\}} categories</cluster\_table>} \\
\end{promptbox}

\noindent \textbf{Prompt 3: Labeling prompt (Step 3 in Section~\ref{sec:taxonomy}).}

\begin{promptbox}
\textbf{Instruction} \\
\vspace{0.5em}
\textbf{Context} \\
- You are given a single conversation summary between a \textbf{User} and an \textbf{AI assistant} that includes \textbf{user-uploaded images}. \\
- Your goal is to identify which \textbf{multimodal capabilities} the user is leveraging, selecting both a \textbf{primary} and (if applicable) a \textbf{second} capability. \\
- You will be provided with a \textbf{reference table} of multimodal capability categories. \\
\vspace{1em}
\textbf{Reference Table} \\
Below is a markdown table of possible categories. Each row defines: \\
- \textbf{id}: a unique index. \\
- \textbf{Label}: the name of a multimodal capability. \\
- \textbf{Description}: a concise explanation of that capability. \\
\vspace{0.5em}
\textcolor{grassgreen}{
\texttt{| id  | Label                    | Description |} \\
\texttt{|------|--------------------------|-------------|} \\
\texttt{| 1    | Recognition              | Identify, classify, and describe objects, scenes, actions, and attributes within images.  |} \\
\texttt{| 2    | Knowledge                | Retrieve, synthesize, and apply factual, commonsense, and contextual information—from both text and visual inputs.  |} \\
\texttt{| 3    | OCR \& Text Extraction  | Detect, read, and interpret text embedded within images/documents via OCR.  |} \\
\texttt{| 4    | Spatial Awareness        | Interpret and reason about positions, relationships, and layouts of objects or text within visual inputs.  |} \\
\texttt{| 5    | Language \& Text Generation    | Produce coherent and context-aware natural-language outputs based on multimodal inputs.  |} \\
\texttt{| 6    | Reasoning                | Apply structured logic and general knowledge to draw inferences, analyze relationships, and solve complex problems.  |} \\
\texttt{| 7    | Code Generation          | Create, modify, and interpret code across languages, including from visual prompts like screenshots or UI elements.  |} \\
\texttt{| 8    | Data Analysis            | Analyze data from charts, graphs, tables, and documents; identify trends and produces structured insights.  |} \\
\texttt{| 9    | Image Editing \& Generation    | Transform or create images via user input—editing, inpainting, style transfer, and generating new visuals.  |} \\
\texttt{| 10   | Document Processing      | Ingest and transform content across file types—summarizing, extracting, reformatting, and converting documents.  |} \\
}
\vspace{1em}
\textbf{Data} \\
- \textbf{"Text"} is a summary of the conversation, including how the user incorporated images as well as their step-by-step workflow with the AI model. \\
Here is the summary: \\
\textcolor{salmonred}{\textbf{\{\{Text\}\}}} \\
\vspace{1em}
\textbf{Task} \\
- Review the conversation and reference table, then \textbf{simultaneously} choose up to three categories: \\
\quad 1. \textbf{Primary Category} \\
\quad\quad - Choose the most specific category (by \texttt{id} and \texttt{Label}) from the reference table that best matches the user's primary multimodal need. \\
\quad\quad - If the user's need does not match any category, output \texttt{-1} for \texttt{id} and \texttt{Undefined} for \texttt{Label}. \\
\quad 2. \textbf{Second Category} \\
\quad\quad - If a second, related multimodal capability is present, choose that as your secondary category (by \texttt{id} and \texttt{Label}). \\
\quad\quad - If no secondary capability applies, output \texttt{-1} and \texttt{Undefined}. \\
\quad 3. \textbf{Explanation} \\
\quad\quad - In no more than \textbf{200} words, explain why the chosen categories (primary and secondary) accurately reflect the multimodal requirements for this task. \\
\quad\quad - Emphasize \textbf{specific details} of the conversation and avoid picking a broad/general category if a more detailed one fits. \\
\vspace{1em}
\textbf{Output Format} \\
Provide your answers exactly between the following XML-style tags (in English): \\
\texttt{<primary-category-id>…</primary-category-id>} \\
\texttt{<primary-category-name>…</primary-category-name>} \\
\texttt{<secondary-category-id>…</secondary-category-id>} \\
\texttt{<secondary-category-name>…</secondary-category-name>} \\
\texttt{<explanation>…</explanation>} \\
\vspace{1em}
- Do not include any other text or tags. \\
- If a category does not apply, use \texttt{-1} for the \texttt{id} and \texttt{Undefined} for the \texttt{Label}. \\
\end{promptbox}

\vspace{1em}
\noindent \textbf{Prompt 4: Developing subcategories (Step 4 in Section~\ref{sec:taxonomy}).}

\begin{promptbox}
category\_name = "
\color{grassgreen}
\texttt{Recognition}
\color{black}" \\
category\_description = "
\color{grassgreen}
\texttt{Visual recognition captures the model's ability to identify, classify, and describe objects, scenes, actions, and attributes within images. This includes tasks such as object detection and counting, scene understanding, color and shape analysis, and domain-specific recognition (e.g. species, brands, artistic styles).}
\color{black}" \\
\end{promptbox}

\begin{promptbox}
\textbf{Instructions} \\
\vspace{0.5em}
- You are given a set of conversation summaries (each a \texttt{"Text"} value) between a \textbf{User} and an \textbf{AI assistant}. \\
- These conversations all include \textbf{user-uploaded images}. \\
- Your goal is to classify what \textbf{multimodal capabilities} each conversation is leveraging. \\
- All conversations fall under one \textbf{parent multimodal category} (e.g., "Recognition"). \\
- Your task is to \textbf{subdivide} these examples into meaningful subcategories based on the parent category's definition. \\

\vspace{1em}
\textbf{Context} \\
- \textbf{Parent category description:} \\
\textcolor{grassgreen}{\texttt{\{\{category\_description\}\}}} \\

\vspace{1em}
\textbf{Input User Conversation Structure} \\
- You are given a markdown table with one row per conversation, structured as follows: \\
  - \texttt{id}: conversation index. \\
  - \texttt{summary}: summarization of the conversation between user and assistant. \\

\vspace{1em}
\textbf{User Conversation Summary Table} \\
\textcolor{salmonred}{\textbf{\{\{Text\}\}}} \\

\vspace{1em}
\textbf{Output Requirements} \\

\textbf{Format} \\
- Output a \textbf{markdown table} where each row defines a subcategory: \\
\quad • \texttt{id}: category index starting from 1. \\
\quad • \texttt{name}: a concise category name (within \texttt{\textcolor{salmonred}{\{\{cluster\_name\_length\}\}}} words). Use either a \emph{noun phrase} or \emph{verb phrase}. \\
\quad • \texttt{description}: a brief explanation (within \texttt{\textcolor{salmonred}{\{\{cluster\_description\_length\}\}}} words) aligned with the parent category. \\

Example format: \\
\texttt{"""markdown} \\
\texttt{| id | name                  | description                              |} \\
\texttt{|----|-----------------------|------------------------------------------|} \\
\texttt{| 1  | Object Identification | Identify specific objects in an image    |} \\
\texttt{"""} \\

- Output must contain \textbf{no more than \texttt{\textcolor{salmonred}{\{\{max\_num\_clusters\}\}}} categories}. \\
- Use \textbf{English only}. \\

\vspace{1em}
\textbf{Quality Guidelines} \\
- No \textbf{overlap} or \textbf{contradiction} among categories. \\
- \texttt{Name} should be concise, specific, and aligned with the category's function. Avoid vague or overly broad terms. \\
- \texttt{Description} should clearly distinguish the category from others. \\
- Ensure \texttt{name} and \texttt{description} are \textbf{consistent}, \textbf{non-redundant}, and \textbf{faithful to the data}. \\
- Output should support high-quality classification for similar future examples. \\

\vspace{1em}
\textbf{Output Tags} \\
Provide your final output between the following XML-style tags: \\
\texttt{<cluster\_table>} your markdown table here \texttt{</cluster\_table>} \\
\end{promptbox}

\section{Robustness Check and Human Validation}
\label{appendix:human_validation}

\subsection{Taxonomy Robustness}

We evaluate whether the induced taxonomy is robust to prompt configurations and generalizes across datasets.

The primary hyperparameter controlling taxonomy induction is \texttt{max\_num\_clusters}, which specifies the maximum number of capability categories. We repeat taxonomy induction using limits of 10 and 20 clusters on the Copilot dataset, running each configuration three times. Note that this parameter specifies an upper bound rather than a fixed number, allowing the LLM to induce fewer categories when appropriate.

To evaluate stability, we canonicalize semantically equivalent labels (e.g., \emph{Visual Recognition} and \emph{Recognition}) and compare consecutive taxonomies using the Levenshtein similarity ratio. Convergence is defined as five consecutive iterations with similarity above 0.85.

The induced taxonomies for each run are shown below.

\paragraph{Copilot dataset (10 clusters, 3 runs).}

\begin{promptbox}
\textbf{Run 1} \\
\color{grassgreen}
\texttt{Recognition; Text Extraction; Image Editing; Code Generation; Spatial Awareness; Reasoning; Knowledge; Language Generation; Data Analysis; Document Handling}
\color{black}
\end{promptbox}

\begin{promptbox}
\textbf{Run 2} \\
\color{grassgreen}
\texttt{Recognition; Knowledge; OCR; Spatial Awareness; Text Generation; Reasoning; Image Editing; Data Analysis; Translation; Coding Assistance}
\color{black}
\end{promptbox}

\begin{promptbox}
\textbf{Run 3} \\
\color{grassgreen}
\texttt{Recognition; Knowledge; OCR; Spatial Awareness; Language Generation; Reasoning; Data Analysis; Image Manipulation; Code Generation}
\color{black}
\end{promptbox}

\paragraph{Copilot dataset (20 clusters, 3 runs).}

\begin{promptbox}
\textbf{Run 1} \\
\color{grassgreen}
\texttt{Recognition; Knowledge; Text Extraction; Spatial Awareness; Language Generation; Reasoning; Image Editing; Code Generation}
\color{black}
\end{promptbox}

\begin{promptbox}
\textbf{Run 2} \\
\color{grassgreen}
\texttt{Recognition; Knowledge; OCR; Spatial Awareness; Language Generation; Reasoning; Image Manipulation; Image Generation; Programming Assistance; Data Analysis}
\color{black}
\end{promptbox}

\begin{promptbox}
\textbf{Run 3} \\
\color{grassgreen}
\texttt{Recognition; Knowledge; Image Analysis; Visual Content Generation; Spatial Awareness; Language Generation; Reasoning; OCR; Document Processing}
\color{black}
\end{promptbox}

We note that across both 10- and 20-cluster settings, the induced taxonomies consistently result in no more than 10 categories. To assess generalizability, we further additionally apply the 10-cluster configuration to the ChatGPT validation dataset.

To assess generalizability, we repeat taxonomy induction on the ChatGPT validation dataset using the same procedure.

\paragraph{ChatGPT dataset (10 clusters, 3 runs).}

\begin{promptbox}
\textbf{Run 1} \\
\color{grassgreen}
\texttt{Image Generation; Image Modification; Visual Recognition; Text Generation; Data Visualization; Knowledge; Spatial Awareness}
\color{black}
\end{promptbox}

\begin{promptbox}
\textbf{Run 2} \\
\color{grassgreen}
\texttt{Image Generation; Image Modification; Recognition; Knowledge; OCR; Spatial Awareness; Language Generation; Reasoning}
\color{black}
\end{promptbox}

\begin{promptbox}
\textbf{Run 3} \\
\color{grassgreen}
\texttt{Recognition; Image Editing; Knowledge; Optical Character Recognition (OCR); Language Generation; Reasoning; Spatial Awareness; Artistic Creation; Data Visualization}
\color{black}
\end{promptbox}

Two researchers independently reviewed all generated labels across runs and datasets and merged semantically equivalent variants into canonical capability categories. Disagreements were resolved through discussion until consensus was reached. The resulting mapping is shown in Table~\ref{tab:label_mapping}. Across all configurations and datasets, the induced taxonomies consistently converged to the same ten capability categories presented in the main paper.

\begin{table*}[t]
\centering
\small
\caption{Mapping of LLM-generated category variants to the final taxonomy. Two researchers manually grouped semantically similar labels into the ten canonical capability categories.}
\label{tab:label_mapping}
\begin{tabular}{p{7cm} p{5cm}}
\toprule
\textbf{Observed Label Variants} & \textbf{Final Label} \\
\midrule
Recognition; Visual Recognition; Image Analysis 
& Recognition \\
\addlinespace[0.4em]
Knowledge 
& Knowledge \\
\addlinespace[0.4em]
OCR; Optical Character Recognition (OCR); Text Extraction 
& OCR \& Text Extraction \\
\addlinespace[0.4em]
Spatial Awareness 
& Spatial Awareness \\
\addlinespace[0.4em]
Language Generation; Text Generation; Translation 
& Language \& Text Generation \\
\addlinespace[0.4em]
Reasoning 
& Reasoning \\
\addlinespace[0.4em]
Code Generation; Coding Assistance; Programming Assistance 
& Code Generation \\
\addlinespace[0.4em]
Data Analysis; Data Visualization 
& Data Analysis \\
\addlinespace[0.4em]
Image Editing; Image Manipulation; Image Modification; Image Generation; Visual Content Generation; Artistic Creation 
& Image Editing \& Generation \\
\addlinespace[0.4em]
Document Handling; Document Processing 
& Document Processing \\
\bottomrule
\end{tabular}
\end{table*}

\subsection{Human Validation of LLM Annotations}

We evaluate whether LLM-generated capability labels agree with human judgments.

\paragraph{Annotation setup.}

Annotators were provided with session summaries together with the definitions of the ten capability categories and independently assigned up to two labels (primary and secondary) to each session.

\paragraph{Data.}

We randomly sampled 200 sessions from the Copilot dataset. Two annotators (one researcher and one graduate student familiar with the taxonomy) independently labeled all sessions.

\paragraph{Evaluation metric.}

Because each session may receive multiple unordered labels, Cohen's $\kappa$ is not directly applicable. Instead, we evaluate agreement using set-level precision, recall, F1 score, and hit rate.

Given two label sets $A$ and $B$,

\[
\mathrm{Precision}(A,B)=\frac{|A\cap B|}{|A|},
\]

\[
\mathrm{Recall}(A,B)=\frac{|A\cap B|}{|B|},
\]

\[
F1(A,B)=
\frac{
2\cdot
\mathrm{Precision}(A,B)
\cdot
\mathrm{Recall}(A,B)
}
{
\mathrm{Precision}(A,B)+
\mathrm{Recall}(A,B)
},
\]

and

\[
\mathrm{Hit}(A,B)=
\begin{cases}
1,&A\cap B\neq\emptyset,\\
0,&\text{otherwise}.
\end{cases}
\]

Hit rate is averaged over all sessions.

We report both human--human agreement and LLM--human agreement in Table~\ref{tab:human_eval}.

\begin{table}[t]
\centering
\small
\caption{Human validation results.}
\label{tab:human_eval}
\begin{tabular}{lccc}
\toprule
\textbf{Metric} &
\shortstack{\textbf{Human}\\\textbf{Agreement}} &
\shortstack{\textbf{LLM vs}\\\textbf{Annotator 1}} &
\shortstack{\textbf{LLM vs}\\\textbf{Annotator 2}} \\
\midrule
Precision & 0.795 & 0.785 & 0.808 \\
Recall    & 0.778 & 0.753 & 0.755 \\
F1 score  & 0.783 & 0.763 & 0.773 \\
Hit rate  & 0.960 & 0.965 & 0.955 \\
\bottomrule
\end{tabular}
\end{table}

Intercoder agreement between the two annotators is high (F1 = 0.783; hit rate = 0.960), indicating strong consistency in human labeling. LLM--human agreement is comparable (F1 = 0.763--0.773), suggesting that the LLM produces capability labels closely aligned with human judgments. Consistent with prior work, F1 scores in this range represent strong agreement for multi-label annotation tasks~\citep{marchal2022establishing}.

\subsection{Eyes-on Validation of the Summarization Pipeline}

The main analysis relies on privacy-preserving \emph{eyes-off} summaries rather than direct access to images. We therefore evaluate whether the summarization step preserves the capability information required for taxonomy induction.

We randomly sample 200 sessions and pair each conversation with its corresponding task and image summaries. Since raw images are inaccessible for privacy reasons, the conversation serves as the closest available proxy to the original interaction.

Two human annotators and \texttt{gpt-4o-mini} independently assign capability labels using (1) the original conversation and (2) the generated summaries. Table~\ref{tab:summary_validation} reports the resulting agreement.

Human agreement remains high between the two conditions (F1 = 0.73; hit rate = 0.91). Likewise, the LLM achieves comparable performance when labeling summaries instead of conversations (F1/hit rate = 0.80/0.96 versus 0.85/0.95).

These results indicate that the summarization step preserves the capability information required for taxonomy induction and dataset annotation while enabling privacy-preserving analysis.

\begin{table}[t]
\centering
\small
\caption{Eyes-on validation of the summarization pipeline. Human agreement measures consistency between labels assigned from conversations and summaries. LLM performance is evaluated against human labels under each condition.}
\label{tab:summary_validation}
\begin{tabular}{lcc}
\toprule
\textbf{Evaluation} & \textbf{F1} & \textbf{Hit Rate} \\
\midrule
Human: Conversation vs. Summary & 0.73 & 0.91 \\
LLM: Conversation & 0.85 & 0.95 \\
LLM: Summary & 0.80 & 0.96 \\
\bottomrule
\end{tabular}
\end{table}

\subsection{Benchmark-Annotation Human Validation}
\label{app:benchmark-annotation-validation}

We validated the \texttt{GPT-5.6-sol} adjudicator used to resolve capability-label disagreements between \texttt{GPT-5.4} and \texttt{Gemini-3.1-pro-preview}. Validation was restricted to the 864 benchmark tasks that both primary annotators identified as involving static images. Among these, 500 (57.9\%) required adjudication because the annotators assigned different capability sets.

\paragraph{Sampling.}
We stratified the 500 adjudicated tasks by the Jaccard similarity of the two annotators' capability sets and sampled 71 tasks, with greater representation of stronger disagreements (Table~\ref{tab:app-sampling}).

\begin{table}[h]
\centering
\caption{Human-validation sample stratified by inter-annotator disagreement.}
\label{tab:app-sampling}
\begin{tabular}{lccc}
\toprule
Bucket & Jaccard similarity & Population & Sampled \\
\midrule
disjoint & $J = 0$ & 11 & 11 \\
low & $0 < J < 0.34$ & 121 & 30 \\
mid & $0.34 \le J \le 0.67$ & 328 & 20 \\
high & $J > 0.67$ & 40 & 10 \\
\midrule
\textbf{Total} & & \textbf{500} & \textbf{71} \\
\bottomrule
\end{tabular}
\end{table}

For each sampled task, a human annotator independently assessed the required taxonomy capabilities and reconciled the assessment with the model annotations to establish a final gold label. We evaluated the adjudicator against these labels using the same agreement metrics as in the main benchmark analysis.

\paragraph{Result.}
Human--adjudicator agreement was macro $\kappa = 0.579$ and mean F1 $= 0.761$, compared with macro $\kappa = 0.679$ and mean F1 $= 0.802$ for \texttt{GPT-5.4}--\texttt{Gemini-3.1-pro-preview} agreement on the full image-task population. 

\newpage
\clearpage
\section{Detailed Taxonomy Description}
\label{appendix:description}

\begin{table}[h]
\caption{Taxonomy of multimodal capabilities with detailed descriptions (up to 200 words per category). This version complements the concise 50-word descriptions shown in Table~\ref{tab:taxonomy} in the main text.}
{\fontsize{8}{9}\selectfont
\begin{tabularx}{\textwidth}{@{}c L{2.5cm} L{10cm}@{}}
\toprule
\textbf{Id} & \textbf{Label} & \textbf{Description} \\
\midrule
1 & Recognition & Visual recognition captures the model's ability to identify, classify, and describe objects, scenes, actions, and attributes within images. This includes tasks such as object detection and counting, scene understanding, color and shape analysis, and domain-specific recognition (e.g. species, brands, artistic styles). \\
\addlinespace[0.3em]
2 & Knowledge & Knowledge covers the model's ability to retrieve, synthesize, and apply factual, commonsense, and contextual information—including social, historical, scientific, and time-sensitive knowledge—from both text and visual inputs to answer questions, provide explanations, and solve complex, multi-step problems across diverse domains. \\
\addlinespace[0.3em]
3 & OCR \& Text Extraction & OCR \& Text extraction captures the model's proficiency in detecting, reading, and interpreting text embedded within images and documents via Optical Character Recognition (OCR). This includes extracting printed or handwritten text and converting it into editable, machine-readable formats. \\
\addlinespace[0.3em]
4 & Spatial Awareness & Spatial awareness captures the model's ability to interpret and reason about the precise positioning, relationships, and interactions of objects and text regions within images or diagrams. This includes understanding layouts, distances, dimensions, and configurations (e.g. maps, flowcharts, geometric figures), enabling tasks such as scene composition analysis, layout design, and spatial context interpretation. \\
\addlinespace[0.3em]
5 & Language \& Text Generation & Language generation captures the model's ability to produce coherent, engaging, and contextually appropriate natural-language outputs according to multimodal inputs. This includes crafting summaries, paraphrases, translations, and creative content (stories, captions, marketing copy); generating structured documents (emails, reports, code comments); and offering detailed explanations—adapting tone, style, and depth to user needs and multimodal inputs. \\
\addlinespace[0.3em]
6 & Reasoning & Reasoning encompasses the model's ability to apply structured, rule-based logic and general world knowledge to draw inferences, analyze relationships, perform cause-and-effect and mathematical reasoning, and solve complex, multi-step problems by synthesizing insights from combined visual and textual inputs. \\
\addlinespace[0.3em]
7 & Code Generation & Code generation captures the model's capacity to create, modify, and optimize programming code across languages based on both textual and visual prompts. This includes generating scripts, functions, classes, and algorithms to meet user specifications; debugging, refactoring, and correcting existing code; interpreting code snippets or UI designs from images; and producing explanatory comments, documentation, or sample implementations for web development, data analysis, automation, and other software tasks. \\
\addlinespace[0.3em]
8 & Data Analysis & Data analysis captures the model's ability to interpret and analyze numerical and textual data from diverse formats—including images, charts, graphs, spreadsheets, and documents—by extracting and summarizing key metrics, performing statistical and arithmetic calculations, identifying trends and relationships, and producing structured outputs (tables, summaries, reports) to generate actionable insights. \\
\addlinespace[0.3em]
9 & Image Editing \& Generation & Image editing \& generation captures the model’s ability to both transform existing images and synthesize new visuals based on user inputs. This includes editing operations—cropping, background removal, color, brightness and contrast adjustments, sharpening, filter and effect application, and object insertion or removal—as well as creative generation tasks like style transfer, upscaling, inpainting, and producing entirely new scenes, illustrations, or designs tailored to specified themes or prompts. \\
\addlinespace[0.3em]
10 & Document Processing & Document processing captures the model’s ability to interpret, extract, and manage structured and unstructured information from documents—including forms, reports, contracts, and PDFs—by leveraging layout understanding, visual-text alignment, and natural language capabilities. Tasks include form parsing, table extraction, document classification, and multi-field entity extraction across varied domains. \\
\bottomrule
\end{tabularx}
\par}
\end{table}

\newpage
\clearpage
\section{Capability Combination Examples}
\label{appendix:combination_examples}

Table~\ref{tab:top_combinations} reports the most frequent capability combinations and representative task examples, highlighting that multimodal interactions are inherently compositional, requiring coordinated perception, reasoning, and generation rather than isolated capabilities.

{\fontsize{8.5}{9}\selectfont
\begin{longtable}{@{}c p{3cm} c p{8cm}@{}}

\caption{Top 10 capability combinations with counts and representative task examples.}
\label{tab:top_combinations} \\

\toprule
\textbf{Id} & \textbf{Capability Combination} & \textbf{Count} & \textbf{Example Task Summaries} \\
\midrule
\endfirsthead

\toprule
\textbf{Id} & \textbf{Capability Combination} & \textbf{Count} & \textbf{Example Task Summaries} \\
\midrule
\endhead

\bottomrule
\endfoot

1 & Knowledge + Language \& Text Generation & 3,563 &
\textit{\textbullet\ Generate a comprehensive set of multiple-choice, short-answer, and explanatory questions for a Software Engineering and Database Design syllabus.} \newline
\textit{\textbullet\ Answer Hindi questions from a 10th-grade CBSE syllabus with detailed explanations of literary concepts and themes.} \newline
\textit{\textbullet\ Create a Web Content Management System (WCMS) project using WordPress by analyzing uploaded materials and producing structured documentation and plans.} \newline
\textit{\textbullet\ Summarize research on continuous carburizing furnaces, including applications, processes, and recent AI-driven advancements.} \newline
\textit{\textbullet\ Complete a mental health case worksheet by structuring patient information and suggesting interventions based on provided clinical details.} \\

\addlinespace[0.4em]
2 & Reasoning + Recognition & 3,437 &
\textit{\textbullet\ Calculate input impedance of electrical circuits from uploaded diagrams by identifying components and applying transformation formulas.} \newline
\textit{\textbullet\ Solve scale drawing and unit conversion problems from worksheet images with step-by-step calculations.} \newline
\textit{\textbullet\ Determine the value of \(k\) in coordinate geometry problems using the distance formula from an image-based question.} \newline
\textit{\textbullet\ Convert an ER diagram (e.g., maritime database schema) into a relational schema with correct primary and foreign keys.} \newline
\textit{\textbullet\ Analyze sound wave diagrams and pendulum setups to compute frequency, pitch, and physical properties.} \\

\addlinespace[0.4em]
3 & Knowledge + Recognition & 3,330 &
\textit{\textbullet\ Develop branding strategies for a YouTube channel using visual references such as thumbnails and gameplay content.} \newline
\textit{\textbullet\ Identify and classify abstract or organic shapes from uploaded images and explain their properties.} \newline
\textit{\textbullet\ Analyze a logo (e.g., ``SUPER DONUTS'') to derive color meanings, typography, and brand guidelines.} \newline
\textit{\textbullet\ Interpret classroom slides on probability and statistics to explain concepts and formulas.} \newline
\textit{\textbullet\ Review patient assessment slides to support care coordination decisions and address social determinants of health.} \\

\addlinespace[0.4em]
4 & Language \& Text Generation + Recognition & 2,291 &
\textit{\textbullet\ Extract and categorize items from an image to produce a structured inventory or assessment report.} \newline
\textit{\textbullet\ Summarize message screenshots to draft a formal Veterans Affairs disability claim statement.} \newline
\textit{\textbullet\ Generate detailed, point-wise descriptions of thumbnail images for content analysis.} \newline
\textit{\textbullet\ Create descriptive alt text for a collage image (e.g., Marvel stickers) highlighting characters and quotes.} \newline
\textit{\textbullet\ Interpret scientific diagrams (e.g., substance behavior under pressure) and explain concepts using simplified analogies.} \\

\addlinespace[0.4em]
5 & Knowledge + Reasoning & 2,132 &
\textit{\textbullet\ Explain healthcare policies (e.g., HIPAA, Patient Bill of Rights) and analyze consent and insurance scenarios.} \newline
\textit{\textbullet\ Solve fraction-based math problems and determine highest powers via LCM and factorization.} \newline
\textit{\textbullet\ Derive physics quantities (e.g., magnetic induction, velocity angles) in electron motion problems.} \newline
\textit{\textbullet\ Explain financial systems, including money supply (M1/M2), banking roles, and interest rate effects.} \newline
\textit{\textbullet\ Evaluate software tools (e.g., Copilot Pro) by analyzing features, integration, and productivity impact.} \\

\addlinespace[0.4em]
6 & Code Generation + Recognition & 1,643 &
\textit{\textbullet\ Debug JavaFX applications by analyzing code snippets and fixing import, annotation, and configuration errors.} \newline
\textit{\textbullet\ Generate Java programs (e.g., employee salary systems) from problem statements provided in images.} \newline
\textit{\textbullet\ Build Angular-based web applications for medical product sales using UI designs and feature requirements.} \newline
\textit{\textbullet\ Follow GitHub workflows by interpreting interface screenshots and executing merge and push operations.} \newline
\textit{\textbullet\ Develop a Fantasy Cricket game in Python using mockup images to guide database, UI, and scoring design.} \\

\addlinespace[0.4em]
7 & Data Analysis + Recognition & 1,150 &
\textit{\textbullet\ Compute financial metrics (e.g., margin, turnover, ROI) from comparative data tables in images.} \newline
\textit{\textbullet\ Recommend appropriate chart types (e.g., line, bubble) for visualizing industrial and risk data.} \newline
\textit{\textbullet\ Extract and reconstruct project folder structures (e.g., Next.js) from directory screenshots.} \newline
\textit{\textbullet\ Generate captions and analytical reports from presentation slides containing performance metrics.} \newline
\textit{\textbullet\ Analyze spending distributions from charts to derive insights (e.g., Pareto patterns, savings opportunities).} \\

\addlinespace[0.4em]
8 & OCR \& Text Extraction + Recognition & 985 &
\textit{\textbullet\ Extract text from images and answer questions based on the extracted content.} \newline
\textit{\textbullet\ Read and extract values from tabular data in images for structured analysis.} \newline
\textit{\textbullet\ Review resumes from uploaded images, providing feedback and identifying errors or improvements.} \newline
\textit{\textbullet\ Answer multiple-choice questions from scanned documents (e.g., networking or Linux exams).} \newline
\textit{\textbullet\ Generate descriptive alt text by extracting and summarizing textual and visual elements from images.} \\

\addlinespace[0.4em]
9 & OCR \& Text Extraction + Language \& Text Generation & 906 &
\textit{\textbullet\ Draft structured emails (e.g., internal announcements for QA sessions) with subject lines.} \newline
\textit{\textbullet\ Generate captions and short-form content for platforms such as YouTube and LinkedIn.} \newline
\textit{\textbullet\ Create optimized product titles for e-commerce listings (e.g., eBay listings under character limits).} \newline
\textit{\textbullet\ Edit and refine project reports to improve tone and contextual relevance.} \newline
\textit{\textbullet\ Compose professional social media posts summarizing events and expressing acknowledgments.} \\

\addlinespace[0.4em]
10 & Code Generation + Reasoning & 859 &
\textit{\textbullet\ Implement Excel logic (e.g., conditional formatting rules based on value ranges) using formulas.} \newline
\textit{\textbullet\ Develop end-to-end Python pipelines for tasks such as customer segmentation.} \newline
\textit{\textbullet\ Solve constraint problems (e.g., 4-Queens) using algorithmic approaches like backtracking.} \newline
\textit{\textbullet\ Design and configure network systems (e.g., IPv4 addressing and subnetting).} \newline
\textit{\textbullet\ Analyze and debug code to verify correctness (e.g., validating algorithm outputs).} \\

\end{longtable}
}

Table~\ref{tab:ood_combinations_long} presents the most frequent capability combinations among OOD samples with representative task examples. These results show that OOD cases are dominated by compositional tasks, particularly those combining recognition with reasoning or code generation, highlighting that multimodal task-space expansion arises from cross-capability interactions.

{\fontsize{8.5}{9}\selectfont
\begin{longtable}{@{}c p{3cm} c p{8.5cm}@{}}

\caption{Top 10 capability combinations among OOD samples with representative task examples.}
\label{tab:ood_combinations_long} \\

\toprule
\textbf{Id} & \textbf{Capability Combination} & \textbf{Count} & \textbf{Example Task Summaries} \\
\midrule
\endfirsthead

\toprule
\textbf{Id} & \textbf{Capability Combination} & \textbf{Count} & \textbf{Example Task Summaries} \\
\midrule
\endhead

\bottomrule
\endfoot

1 & Recognition + Reasoning & 512 &
\textit{\textbullet\ Solve a map coloring problem using forward checking with three colors (Red, Green, Blue), with step-by-step illustration and summary.} \newline
\textit{\textbullet\ Determine the manometer reading for a conical vessel filled with water using measurements from an uploaded image.} \newline
\textit{\textbullet\ Identify congruent rectangles from an image by comparing dimensions and geometric properties.} \newline
\textit{\textbullet\ Solve for variables \(x\) and \(y\) in a geometry problem involving parallel lines and transversals using angle relationships.} \newline
\textit{\textbullet\ Analyze a signal processing problem involving the unit step function, including sketches and sampling instructions.} \\
\addlinespace[0.4em]
2 & Code Generation + Recognition & 256 &
\textit{\textbullet\ Convert a code snippet from an uploaded image into Python code and provide installation instructions for required libraries.} \newline
\textit{\textbullet\ Generate an HTML layout for a web page based on an uploaded UI design image, including styling and structure.} \newline
\textit{\textbullet\ Modify a navbar by removing dropdown arrows, enabling hover interactions, and adding profile/cart icons using Bootstrap.} \newline
\textit{\textbullet\ Generate SQL code from a table image and normalize it to third normal form with explanations.} \newline
\textit{\textbullet\ Create a React Native screen with a centered logo and custom styling based on design requirements.} \\
\addlinespace[0.4em]
3 & Knowledge + Language \& Text Generation & 220 &
\textit{\textbullet\ Explore methods for transcribing audio files using tools such as Microsoft Word, Flixier, and manual transcription.} \newline
\textit{\textbullet\ Draft a tailored cover letter for a Signallers Manager role using detailed CV information and job requirements.} \newline
\textit{\textbullet\ Create structured notes and summaries for a Year 12 Physics assignment on gravitational waves.} \newline
\textit{\textbullet\ Summarize an academic article on test validity and integrative evaluation frameworks.} \newline
\textit{\textbullet\ Improve a resume by incorporating updated skills, achievements, and industry-relevant keywords.} \\
\addlinespace[0.4em]
4 & Recognition + Knowledge & 196 &
\textit{\textbullet\ Diagnose a \texttt{terraform init} error from a terminal screenshot and suggest installation and PATH fixes.} \newline
\textit{\textbullet\ Analyze SEM images to describe particle morphology and size for materials science applications.} \newline
\textit{\textbullet\ Explain Enterprise JavaBeans concepts from an uploaded architecture diagram.} \newline
\textit{\textbullet\ Resolve AWS S3 permission errors by analyzing screenshot messages and suggesting fixes.} \newline
\textit{\textbullet\ Identify font families using image-based font recognition tools.} \\
\addlinespace[0.4em]
5 & Reasoning + Knowledge & 194 &
\textit{\textbullet\ Analyze circuits to compute voltage, current, and power balance using symbolic variables.} \newline
\textit{\textbullet\ Compute error in PV emulator circuits using electrical parameters and I-V characteristics.} \newline
\textit{\textbullet\ Determine continuity conditions for piecewise functions and verify limits.} \newline
\textit{\textbullet\ Evaluate polynomial functions to determine end behavior based on leading terms.} \newline
\textit{\textbullet\ Calculate terminal velocity using particle and fluid properties with detailed steps.} \\
\addlinespace[0.4em]
6 & Code Generation + Reasoning & 184 &
\textit{\textbullet\ Debug code snippets by correcting syntax errors and improving clarity (e.g., printf formatting).} \newline
\textit{\textbullet\ Improve website CSS for responsiveness and eliminate layout overflow issues.} \newline
\textit{\textbullet\ Restructure DataFrames by extracting timestamps and pivoting data using pandas.} \newline
\textit{\textbullet\ Implement chessboard logic to determine if coordinates share the same color square.} \newline
\textit{\textbullet\ Develop Java classes and validation logic, including regex-based checks and test cases.} \\
\addlinespace[0.4em]
7 & Knowledge + Reasoning & 173 &
\textit{\textbullet\ Record accounting transactions, including inventory adjustments and journal entries.} \newline
\textit{\textbullet\ Apply normal approximation to the binomial distribution with step-by-step explanations.} \newline
\textit{\textbullet\ Diagnose GCC compiler errors and resolve PATH configuration issues.} \newline
\textit{\textbullet\ Analyze international legal frameworks for human rights protection in war zones.} \newline
\textit{\textbullet\ Provide recovery steps for BitLocker-encrypted data and discuss limitations without keys.} \\
\addlinespace[0.4em]
8 & Reasoning + \texttt{None} & 159 &
\textit{\textbullet\ Compute average shear stress in beam pins using static equilibrium and cross-sectional analysis.} \newline
\textit{\textbullet\ Calculate ramp length using height and angle with trigonometric functions.} \newline
\textit{\textbullet\ Solve free-fall physics problems using kinematic equations.} \newline
\textit{\textbullet\ Compare growth rates of functions \(n \log n\) and \(n \sqrt{n}\).} \newline
\textit{\textbullet\ Compute acceleration under electric fields using force and mass relationships.} \\
\addlinespace[0.4em]
9 & OCR \& Text Extraction + Recognition & 157 &
\textit{\textbullet\ Extract and format employment details from certification documents.} \newline
\textit{\textbullet\ Transcribe Hindi letters from images into editable text.} \newline
\textit{\textbullet\ Extract and summarize legal documents such as power of attorney forms.} \newline
\textit{\textbullet\ Transcribe and summarize academic discussions from image content.} \newline
\textit{\textbullet\ Identify and correct capitalization errors in text using extracted content.} \\
\addlinespace[0.4em]
10 & Recognition + Language \& Text Generation & 138 &
\textit{\textbullet\ Generate formatted tables based on image templates of structured data.} \newline
\textit{\textbullet\ Create product descriptions for jewelry images highlighting brand values and features.} \newline
\textit{\textbullet\ Compute geometric areas from diagrams and present results in equation form.} \newline
\textit{\textbullet\ Generate captions for timetable images and contextual explanations.} \newline
\textit{\textbullet\ Produce SEO-optimized alt text for product images.} \\

\end{longtable}
}